\documentclass{aa}

\def\kpc{{\rm kpc}}
\def\kms{{\rm km\,s^{-1}}}

\def\dR{$\partial V_\phi / \partial R$}
\def\dphi{$\partial V_\phi / \partial \phi$}

\def\kmskpc{\,kms$^{-1}$kpc$^{-1}$}
\def\kmsdeg{\,kms$^{-1}$deg$^{-1}$}

\newcommand{\mycomment}[1]{}

\let\oldsim\sim 
\renewcommand{\sim}{{\oldsim}}

\usepackage[normalem]{ulem}
\usepackage{graphicx}
\usepackage{txfonts}
\usepackage{amsmath}
\usepackage{yhmath}
\usepackage{algorithm,algorithmic}
\usepackage{xcolor}
\usepackage{bm}
\usepackage{multirow}
\usepackage{svg}
\usepackage{makecell}

\defcitealias{Ramos_2018}{R18}
\defcitealias{Bernet2022}{B22}

\begin{document} 

   \title{Radial and azimuthal gradients of the moving groups in Gaia DR3: The slow/fast bar degeneracy problem}

   \author{M. Bernet \inst{1,2,3}
            \and P. Ramos \inst{4}
            \and T. Antoja \inst{1,2,3}
            \and G. Monari \inst{5}
            \and B. Famaey \inst{5}
          }
    \institute{Departament de Física Qu\`antica i Astrof\'isica (FQA), Universitat de Barcelona (UB),  C Mart\'i i Franqu\`es, 1, 08028 Barcelona, Spain
           \email{mbernet@fqa.ub.edu}
    \and{Institut de Ci\`encies del Cosmos (ICCUB), Universitat de Barcelona (UB), C Mart\'i i Franqu\`es, 1, 08028 Barcelona, Spain}
    \and{Institut d'Estudis Espacials de Catalunya (IEEC), Edifici RDIT, Campus UPC, 08860 Castelldefels (Barcelona), Spain}
    \and{National Astronomical Observatory of Japan, Mitaka-shi, Tokyo 181-8588, Japan}
    \and{Universit{\'e} de Strasbourg, CNRS, Observatoire astronomique de Strasbourg, 11 rue de l’Universit{\'e}, 67000 Strasbourg, France}\\
    }

   \date{Received YYY; accepted XXX}

 
  \abstract
  {The structure and dynamics of the central bar of the Milky Way (MW) are still under debate whilst being fundamental ingredients for the evolution of our Galaxy. The recent Gaia DR3 offers an unprecedented detailed view of the 6D phase-space of the MW, allowing for a better understanding of the complex imprints of the bar on phase-space.}
  {We aim to identify and characterise the dynamical moving groups across the MW disc, and use their large-scale distribution to help constrain the properties of the Galactic bar.}
  {We used 1D wavelet transforms of the azimuthal velocity ($V_\phi$) distribution in bins of radial velocity to robustly detect the kinematic substructure in the Gaia DR3 catalogue. We then connected these structures across the disc to measure the azimuthal ($\phi$) and radial ($R$) gradients of $V_\phi$ of the moving groups. We simulated thousands of perturbed distribution functions using Backwards Integration, sweeping a large portion of parameter space of feasible Galaxy models that include a bar, to compare them with the data and to explore and quantify the degeneracies.}
  {The radial gradient of the Hercules moving group ($\partial V_\phi/\partial R$ = 28.1$\pm$2.8 km$\,$s$^{-1}\,$kpc$^{-1}$) cannot be reproduced by our simple models of the Galaxy which show much larger slopes both for a fast and a slow bar. This suggests the need for more complex dynamics (e.g. a different bar potential, spiral arms, a slowing bar, a complex circular velocity curve, external perturbations, etc.). We measure an azimuthal gradient for Hercules of $\partial V_\phi/\partial \phi$ = -0.63$\pm$0.13$\,$km$\,$s$^{-1}$deg$^{-1}$ and find that it is compatible with both the slow and fast bar models. Our analysis points out that using this type of analysis at least two moving groups are needed to start breaking the degeneracies.}
  {We conclude that it is not sufficient for a model to replicate the local velocity distribution; it must also capture its larger-scale variations. The accurate quantification of the gradients, especially in the azimuthal direction, will be key for the understanding of the dynamics governing the disc.}
   

   \maketitle

%

%
%

\section{Introduction}


The phase-space distribution function (DF) of stars in the Milky Way (MW) is a key element in understanding the structure and history of our Galaxy.
A precise characterisation of this 6D phase-space (position and velocity $R,\phi,Z,V_R,V_\phi,V_Z$) of the MW has been made possible by the second and third \emph{Gaia} data releases \citep[\emph{Gaia} DR2 and DR3,][]{dr2, dr3}.
Studying this \emph{Gaia} data in various projections reveals the complex substructure within it; ridges in $R-V_\phi$ \citep{antoja2018dynamically,kawata2018radial,fragkoudi2019ridges}, a wave in $L_Z-V_R$ \citep[$L_Z = R V_\phi$, ][]{friske2019wave,antoja2022wave}, a bimodality in $L_Z - V_Z$ \citep{antoja2021anticentre,mcmillan2022}, the phase spiral in $Z-V_Z$ \citep{antoja2018dynamically, antoja2023spiral, hunt2021spiral, hunt2022spiral, laporte2019sag}, and thin arches in $V_R-V_\phi$ \citep{katz2018dynamics,ramos2018}.

Out of all of these projections, the $V_R-V_\phi$ distribution close to the solar neighbourhood (SN) is historically the most studied one \citep{stromberg1946,eggen1965movingroups}. It presents a complex configuration of overdensities --usually called moving groups-- shaped as thin arches \citep{dehnen1998movinggroups,famaey2005mg}. Some of the moving groups, and also the ridges and the $L_Z-V_R$ wave, have been related to the orbital resonances of the bar and spiral arms of the Galaxy \citep{kalnajs1991pattern,raboud1998resonances,Dehnen2000,antoja2011understanding,hunt2018outer,Hunt2018hercules,hunt2018transient,hunt2019signatures,monari2019signatures,Bernet2022} and/or attributed to ongoing  phase-mixing related to external perturbations \citep{minchev2009milky,gomez2012signatures,antoja2018dynamically,ramos2018,hunt2018transient,khanna2019galah,laporte2019sag,laporte2020ages}.
Despite using numerous analytical and numeric approaches to understand the structure of the phase space, we still face a significant challenge in explaining its origin in detail.

In particular, the origin of the Hercules moving group has been a subject of extensive debate. For the past two decades, it has been suggested that this moving group is connected to the resonant interactions between local stars and the central Galactic bar 
\citep[e.g.,][]{Dehnen2000,antoja2014constraints}. According to this hypothesis, if the Sun is located just beyond the Outer Lindblad Resonance (OLR) of the bar, a Hercules-like overdensity can be generated, thus implying a bar pattern speed of $\Omega_b\,\sim\,55$\,\kmskpc \citep[the \textit{short/fast bar} scenario,][]{minchev2007fastbar,chakrabarty2007phasespace,monari2017herculesfast}. However, recent studies of the gas \citep{sormani2015gasbar} and stellar kinematics \citep{portail2017,sanders2019patternspeed} in the inner Galaxy point to a pattern speed of $\Omega_b\,\sim\,40$\,\kmskpc (the \textit{long/slow bar} scenario). Following this new evidence, \citet{perezvillegas2017hercules} and \citet{monari2019signatures} proved that orbits trapped in the co-rotation (CR) resonance of a long/slow bar can generate a Hercules-like overdensity in the local velocity space, although less pronounced than the one produced by the OLR
\citep{Monari2017resonanttrapping,Binney2018orbits,Hunt2018hercules,fragkoudi2019ridges}. However, the combination of spiral arms and a bar \citep{hunt2018transient}, or a decelerating bar \citep{chiba2021decelerating,chiba2021decelerating_ring} can create stronger Hercules-like overdensities even for a slow bar.

Indirectly measuring the pattern speed of the bar requires reproducing the moving groups at the SN through the bar resonances, something that has been mostly done qualitatively (e.g., \citealt{Dehnen2000,perezvillegas2017hercules,hunt2018outer,trick2021action,trick2022angle}, but see \citealt{Asano2022,clarke2022slow}).
As the available data improved, first with RAVE \citep{steinmetz2006rave} and later on with \emph{Gaia}, measurements of the pattern speed have been improved by including other regions of the galactic disk \citep{antoja2014constraints,monari2014largescale,monari2019signatures}.
However, as explained above, there are a few combinations of pattern speeds and resonances that produce a similar local velocity distribution, i.e. degenerated solutions. 
Studying the position of the moving groups across the disc might have the potential to break this degeneracy.

In \citeauthor{Bernet2022} (\citeyear{Bernet2022}, hereafter \citetalias{Bernet2022}) we presented a robust, large-scale characterisation of the moving groups across the MW disc. 
Our analysis revealed that the moving groups exhibit complex spatial changes, deviating from the expected lines of constant angular momenta along the radial direction and showing clear non-axisymmetries in azimuth. In particular, our study confirmed the azimuthal gradient of the Hercules moving group measured in \citet{monari2019hercules}. In their work, they favour the slow bar scenario using this measurement, as well as the significant azimuthal slope of the Horn, which is another moving groups at negative $V_R$ \citep{Monari2013}. Based on the predictions they obtained from perturbation theory, the gradient of $L_z$ with azimuth for a slow-bar Horn (created by the 6:1 resonance) should be non-zero, while the fast-bar Horn (created by OLR) should have a very small azimuthal gradient.

The goal of this study is to quantitatively characterise the moving groups in the phase space using the new \emph{Gaia} DR3 data and improving on the methodology from \citetalias{Bernet2022} by including, in addition to $V_R$ and $V_\phi$, a novel set of measurables: the gradients of the moving groups at the SN in the radial (\dR) and azimuthal (\dphi) directions. The characterisation of the spatial variations of these groups allows for meaningful comparisons with theoretical models.
As a particular case, we investigate the implications of kinematic substructures on the pattern speed of the MW bar, comparing the gradients obtained from the data with those of models with a simple bar, using backward integration simulations. We find that some measurables are compatible with both the fast and the slow bar models while others are incompatible with either.
We also examine whether the gradients in the models truly differ between a slow and a fast bar scenario, and determine the minimum number of observables required to break the degeneracies. We observe a large disagreement between the measured gradients in the data and the simulations. This disagreement leads us to the conclusion that simple models are insufficient to reproduce the current observations.

This paper is organized as follows. In Section \ref{sect:DR3_analysis} we describe the data selection, summarize the methodology presented in \citetalias{Bernet2022}, and explain the computation of the gradients. In Section \ref{sect:simulations}, we introduce the simulations used in this analysis and how we apply our methodology to them. Section \ref{sect:results_1} describes the kinematic substructure observed in the data and the models, and compare the different trends shown by the moving groups and overdensities. In Section \ref{sect:results_2}, we sweep the pattern speed $\Omega_b$ and the slope of the rotation curve $\beta$ in the models and compare the obtained measurables with the data. In Section \ref{sect:azimuth_time_simil} we study the change of the overdensities in time and azimuth in the models to understand the properties of the different structures. The implications of these results are discussed in Section \ref{sect:discussion}. Finally, in Section \ref{sect:conclusions} we summarise our results and list the main conclusions of this work.

%
%

\section{Data and methodology}\label{sect:DR3_analysis}

\subsection{Data}

We used data from the \emph{Gaia} Data Release 3 \citep[\emph{Gaia} DR3,][]{dr3}. From it, we selected the approximately 34 million stars with position, proper motion, parallax, and line-of-sight velocity, and used the StarHorse \citep{Anders2022} distances. We applied an astrometric quality selection ($\mathtt{RUWE}<1.4$), a selection in parallax quality ($\mathtt{parallax\_over\_error}>5$), and a selection of non-spurious solutions \citep[$\mathtt{fidelity\_v2}>0.5$,][]{rybizki2022spurious}. As recommended for this sample, we corrected the line-of-sight velocity for stars with $\mathtt{grvs\_mag} \ge 11$ and $\mathtt{rv\_template\_teff} <8500$\,K using Eq. 5 in \citet{Katz2022}. \citet{Blomme2023} discussed the need for another correction for stars with $8500 \le \mathtt{rv\_template\_teff} \le 14,500$\,K and $6 \le \mathtt{grvs\_mag} \le 12$. However, after the correction, a residual bias of a few $\kms$ remained. Since these stars are rare, we only kept stars with $\mathtt{rv\_template\_teff}<8500$\,K in our sample. Our final sample has 25,397,569 stars.

We transformed the observables into cylindrical phase space coordinates using $R_0=8.277\,\kpc$  \citep{Gravity2022}\footnote{\citet{clarke2022slow} used an alternative measure of $R_0$, the centre of the bulge. In their study they found a great agreement between this measurement and the distance to Sag A*, confirming the hypothesis that Sag A* is at rest at the centre of the bulge.}, $Z_\odot = 0.0208\,\kpc$ \citep{Bennett2019}, and $U_\odot=9.3$, $V_\odot+V_c(R_0)=251.5$, and $W_\odot=8.59\,\kms$ from the proper motion of SagA* \citep{Reid2020} and its radial velocity \citep{Gravity2022}. The reference system is right-handed with $\phi$ oriented contrary to the disc rotation and thus $V_\phi<0$ for most stars.  We set the origin at the Sun's azimuth ($\phi_\odot=0$).

\subsection{Methodology}\label{sect:methodology}

In \citetalias{Bernet2022}, we presented a novel method to extract large kinematic substructures from a dataset of stellar positions and velocities. 
The first step of the method is to partition the data into small spatial volumes $(\Delta R, \Delta \phi, \Delta Z)$ and run a Wavelet Transform (WT) on the velocity distribution $(V_R, V_\phi)$ of each individual volume to detect the overdensities. These overdensities --the moving groups-- are known to form thin arches elongated around large ranges of $V_R$, with a slight variation in $V_\phi$ \citep{ramos2018}. 
Because of this geometry, in the second step, we slice each $(V_R, V_\phi)$ diagram in $\Delta V_R$ bins, and run a 1D WT in each $V_\phi$ histogram. The peaks obtained from these 1D WT are connected through the configuration space to form global substructures using a modification of the Breadth-First Search \citep[BFS,][]{moore1959bfs} algorithm from graph theory. This method has proven to be effective in detecting large kinematic substructures in the \emph{Gaia} EDR3 data and test particle simulations. For more details, we refer to the original paper.

We modified a few parameters of the method with respect to the execution in \citetalias{Bernet2022}. Firstly, we used a larger scale of the wavelet, from $2\,\kms$ in \citetalias{Bernet2022} to $6.25\,\kms$ here. The main goal of the wavelet size modification was to robustly detect the large-scale substructure in regions far from the Sun, where the statistical significance of the moving groups decreases. Secondly, we increased the resolution of the grids used in the configuration and velocity spaces. This reduced the degeneracies in the linking stage of the methodology and allowed for a more precise computation of the gradients (\dR{} and \dphi).

The execution was run in the Cloud using resources granted by the Open Clouds for Research Environments (OCRE) from the European Union. In \citetalias{Bernet2022} we swept the entire disc volume, but in this analysis we focused on the radial ($\phi=0$\,deg, $Z=0$\,kpc) and azimuthal ($R = 8.277$\,kpc, $Z=0$\,kpc) directions. Our new binning to compute the positions of the moving groups was:
\begin{itemize}
    \item Radial direction $(R)$: $[5,14]$ kpc in steps of $0.01$ kpc,\\ $\Delta R = \pm0.24$ kpc around each centre;
    \item Azimuthal direction $(\phi)$: $[-50,50]$ deg in steps of $0.1$ deg,\\ $\Delta\phi = \pm2.4$ deg around each centre;
    \item Vertical direction $(Z)$: Not swept, \\ $\Delta Z = \pm0.24$\,kpc around $Z=0$\,kpc;
    \item Radial velocity $(V_R)$: $[-120,120]\,\kms$ in steps of $2\,\kms$,\\$\Delta V_{R} = \pm 15\,\kms$ around each centre.
\end{itemize}

The possible biases introduced by a wrong distance estimation or the size of the bins were discussed in \citetalias{Bernet2022}. We tested the bias of the bin size using a smaller bin, and estimated the impact to be below $2\kms$, well below the WT size. Since the coordinate transformations and binning were very similar here, we refer to those analyses.

After the execution, we obtained independent detections of each moving group at each small spatial volume, and their link across the space in the form of large-scale ridge-like structures (Fig.\,\ref{fig:vphi_r_phi}), which we analyse later on. By construction, each detection corresponds to a single $V_R$, and each moving group is in turn assigned a set of these detections that span the entirety of its arch in velocity space. 
In the rest of the paper, the term \emph{structure} refers to the grouping for a single $V_R$ in the entire configuration space, either in the data or the simulations. \emph{Moving group} refers to a set of structures covering an entire arch in velocity space in the data, and \emph{overdensity} refers to the equivalent of a moving group in the simulations.

We want to quantify how the structures change across the MW disc. However, visualising how tens of structures evolve along 2 dimensions all at once is challenging. To reduce the dimensionality of the problem, we computed the radial (\dR) and azimuthal (\dphi) gradients of each structure at the SN.
The computation of gradients is numerically unstable, small errors in the input data propagate throughout the computation, leading potentially to big errors. To produce a robust estimation of the gradient, we fitted a parabola to the structures in the radial direction and a straight line in the azimuthal direction (dashed lines in the top plots in Fig.\,\ref{fig:vphi_r_phi}), and computed the analytical derivative of these fittings at the SN. We can then plot these gradients in the SN (left panels in Fig.\,\ref{fig:DR3_gradients}). We propagated the uncertainties in $V_\phi$ for each measurement (WT size, $\sigma = 6.25\,\kms$) using Monte Carlo sampling. The uncertainties in the measurements of the gradients in the main moving groups (Hat, Sirius, Hyades, Horn, and Hercules) are below  $1$\,\kmskpc{ }in \dR, and below $0.05$\,\kmsdeg{ }in \dphi.

%
%
 
\section{Simulations}\label{sect:simulations}

The \emph{Gaia} data is unprecedented in its quality and quantity. As we explained in the previous section, it even allows us to compute the spatial gradients of the velocity structures. Running realistic simulations that match the quality of the data is complex and expensive. For this reason, we used the simpler, but many times faster, backwards integration (BI) method that we explain in Sect. \ref{sect:BI_setup} to explore exhaustively the parameter space of different models. In Sect. \ref{sect:BI_arch_detection} we describe the specific method of detection of structures in the simulations.

\subsection{Setup and bar potential}\label{sect:BI_setup}

The BI technique (\citealt{vauterin1997bi,Dehnen2000}, also \citealt{hunt2018outer} for a more recent example) allows us to approximate the response of the DF to a bar perturbation by integrating the orbit of stars, all starting at a certain point $(R,\phi)$ in configuration space, but with different velocities on a grid in the $(V_R, V_\phi)$ plane.
According to the collisionless Boltzmann equation, the density of the DF inside an infinitesimal phase-space volume remains constant. However, the DF can only be measured in finite volumes.
The BI technique assumes that the mean value within the local volume is the same as its central value, regardless of how the volume is deformed along the orbital evolution. This means that we trace back a single orbit at the central point of each volume to compute the DF before the perturbation. It is important to note, however, that these local volumes undergo significant stretching and kneading, potentially challenging the assumption that the mean is equal to the central value. In other words, the measurable DF, also sometimes called the coarse-grained DF, does not obey the collisionless Boltzmann equation. Moreover, when smoothly switching on the perturbation amplitude up to its plateau, the separatrixes will move in time whilst virtually always involving non-adiabatic behaviour on their surface, implying a dependence on the switch-on choice. Nevertheless, the assumptions made here are still extremely useful, especially for studying the dynamical effect of the perturbation for relatively few dynamical times.

In practice, we follow the model presented in \citet{Dehnen2000}.
To validate our approach, we refer to existing literature: \citet{Monari2017resonanttrapping} modified the setup proposed by \citet{Dehnen2000} to ensure complete phase-mixing and calculated the present-day DF using the pendulum approximation. Additionally, \citet{trick2021action} ran forward test particle simulations in an identical setup. Both studies produced results in agreement with the BI technique, affirming the reliability of our computations and the low impact of the assumptions presented in the previous paragraph.

We use the \citet{dehnen1999df} distribution function to model the stellar disc before the bar formation
\begin{equation}
    f(E,L) \propto \frac{\Sigma(R_E)}{\sigma_R^2(R_E)}\exp\Bigg[\frac{\Omega(R_E)(L-L_c(E))}{\sigma_R^2(R_E)}\Bigg]
\end{equation}
where $R_E$, $\Omega(R_E)$, and $L_c(E)$ are radius, circular frequency, and angular momentum of a circular orbit with energy E. The analytical form of the DF allows for an efficient computation. $\Sigma(R)$ and $\sigma_R(R)$ are defined as
\begin{eqnarray}
    \Sigma(R) &=& \Sigma_0\exp(-R/R_s), \\
    \sigma_R(R) &=& \sigma_0\exp(-R/R_\sigma).
\end{eqnarray}
where the values of the constants $\Sigma_0$, $\sigma_0$, $R_s$, and $R_\sigma$ (Tab.\,\ref{tab:simulation_params}) are the same as in \citet{Dehnen2000}. This DF is used extensively in the literature \citep[e.g.][]{Hunt2018hercules,monari2019signatures}, providing an initial setup consistent with previous analysis. Possible bias induced by the selection of the DF parameters are discussed later on.

The axisymmetric potential is assumed to be such that the circular velocity is given by
\begin{equation}
    v_c (R) = v_0 \big(R/R_0\big)^\beta
\end{equation}
\noindent where $v_0$ is the circular velocity at the Solar radius $R_0$, and $\beta$ characterises the shape of the rotation curve.

Also following \citet{Dehnen2000}, we model the bar potential as a simple quadrupole
\begin{equation}
\Phi_b(R,\phi)=A_b(t)\cos(2(\phi-\phi_{b}-\Omega_b t))\times
\left\{ \begin{array}{ll} -(R_b /R)^3, & \mathrm{for}\ R \geq R_b,\\ (R/R_b)^3 - 2, & \mathrm{for}\ R \leq R_b, \end{array}
\right.
\end{equation}
\noindent where $\Omega_b$ is the pattern speed of the bar, $\phi_b$ is angle of its long axis and $R_b$ its size, which we fix to be $80\%$ of the corotation radius\footnote{This value is the same used by \cite{Dehnen2000} and is close to the reference value of $\mathcal{R}=\frac{R_{CR}}{R_{bar}}=1.2$ (\citealt{Athanassoula1992}, but see \citealt{Font2017}).}, given by
\begin{equation}
    R_{CR}\equiv R_0\big(\Omega_b / \Omega_0 \big)^{1/(\beta-1)},
\end{equation}
where $\Omega_0\equiv v_0/R_0$ denotes the circular frequency at the Sun.

The bar amplitude is initialized at $A_b(t) = 0$ for $t=0$, increases smoothly for $t<t_1$ as
\begin{equation}
    A_b(t) = A_f\biggl[\frac{3}{16}\xi^5-\frac{5}{8}\xi^3+\frac{15}{16}\xi+\frac{1}{2}\biggr], \quad \xi=2\frac{t}{t_1}-1,
\end{equation}
\noindent and remains at a fixed $A_f$ for $t>t_1$ until the end of the simulation at $t = t_2$. We express $t$ in units of the bar rotation period $T_b \equiv 2\pi\Omega_b$. Finally, the amplitude constant is defined as 
\begin{equation}
    A_f = \alpha v_0^{2/3} \Bigg(\frac{R_0}{R_b}\Bigg)^3,
\end{equation}
\noindent which depends on the dimensionless bar strength $\alpha$, defined as the ratio of the radial forces from the bar's quadrupole and the axisymmetric power-law background at $R_0$ along the major axis of the bar \citep{Dehnen2000}.

\begin{table}[]
\renewcommand{\arraystretch}{1.3}
\caption{Parameters of the fiducial models.}
\begin{tabular}{cccc}
\hline \hline
Symbol & Units & SBM & FBM  \\ \hline
$\Omega_b$   & \kmskpc  & 39   & 56   \\
$t_1$        & $Gyr$    & $2T_b = 0.32$   & $2T_b = 0.22$   \\
$t_2$        & $Gyr$    & $4T_b = 0.64$   & $4T_b = 0.44$   \\ \hline
$R_0$        & kpc      & \multicolumn{2}{c}{8}    \\
$v_0$        & $\kms$   & \multicolumn{2}{c}{220}  \\
$\alpha$     &          & \multicolumn{2}{c}{0.01} \\
$\beta$      &          & \multicolumn{2}{c}{0}    \\ 
$\Sigma_0$   & kms$^{-1}$ & \multicolumn{2}{c}{1}  \\ 
$\sigma_0$   & kms$^{-1}$ & \multicolumn{2}{c}{48} \\ 
$R_s$        & kpc      & \multicolumn{2}{c}{2.64} \\ 
$R_\sigma$   & kpc      & \multicolumn{2}{c}{8}    \\ \hline
\end{tabular}
\label{tab:simulation_params}
\renewcommand{\arraystretch}{1}
\end{table}

We present two fiducial models, which aim to reproduce the velocity distribution of the SN in a MW with a fast bar and a slow bar. In the Slow Bar Model (SBM) we included a bar with a pattern speed of $\Omega_b = 39$\,\kmskpc. The Fast Bar Model (FBM) has a pattern speed of $\Omega_b = 56$\,\kmskpc. The Sun is at $R_\odot = 8$\,kpc, and the bar is at $\phi_{b} = -30$\,deg with respect to the Sun \citep{bland-hawthorn2016structure}.
The parameters of the fiducial simulations can be found in Table \ref{tab:simulation_params}.
The orbit integration was performed using AGAMA \citep{vasiliev2019agama}.

For consistency, we tested the effect of a significant modification of the parameters of the DF. We increased individually the fiducial values of $\Sigma_0$, $\sigma_0$, $R_s$, and $R_\sigma$ by a $25\%$ and assessed that the change on the position of the overdensities is below $2\kms$. In addition, we tested the relative error in the gradients with the same increase of $25\%$. The relative error in radial gradients is below $2\%$, while the relative error in azimuthal gradients is below $5\%$. These shifts are even less significant on higher order resonances. With this analysis, we show that the measurements are robust to significant variations of the input parameters.  We remark that the local velocity dispersion of the used model ($\sigma_R = 17.6\,\kms$) would correspond to a relatively younger population of stars of the MW \citep[$1-2$\,Gyr,][]{robin2022vdispersion}. We tested the measurables obtained with a larger dispersion ($\sigma_R = 35\,\kms$) which would correspond to an older population \citep[$6-7$\,Gyr, ][]{robin2022vdispersion} and the variations are well within the errors estimated in Table \,\ref{tab:slopes}.

Finally, we computed the gradients with a \citet{McMillan2017} potential, a \citet{Ferrers1877} bar and a Quasi-isothermal DF \citep{binney2011quasiisothermal}, and observed an error of $5\%$ in radial gradient, and $20\%$ in azimuthal gradient. Therefore, we conclude that the choice of DF has an impact on the final models, but the obtained measurables are still dominated by the global physics of the resonances.

We also note that the used models are 2D for simplicity, ignoring the contribution of the vertical motion. It is known that the position of the moving groups depends on the vertical component \citepalias{Bernet2022}. However, the studied sample is very dominated by stars close to the plane, and we tested that different vertical sizes in the selection induce biases below $1\,\kms$ in the positions of the moving groups. Therefore, we conclude that the vertical substructure will not bias our measurements, thus allowing for a fair comparison with 2D models.

\subsection{Arch detection}\label{sect:BI_arch_detection}

In these simulations, we applied the methodology described in Sect.\,\ref{sect:methodology}. The first step of this methodology is to detect overdensities in the velocity distribution. For the data, the WT has proven to be optimal for this task \citep[e.g.][]{Chereul1998,Skuljan1999,antoja2008moving,Kushniruk2017,ramos2018,lucchini2023a}. However, for the simulated velocity distributions, we find that in the case of the SBM the detections with the WT are not robust enough. This is because the overdensity caused by CR is not as prominent as when observed in the data or the FBM \citep{Binney2018orbits, Hunt2018hercules}. Luckily, the velocity distributions produced with BI have potentially ``infinite'' resolution\footnote{In practice, to have reasonable computing costs, we limit our velocity grid to a resolution of $0.2$~$\kms$, which is of the order of the velocity uncertainties at the SN.}, and can be considered twice-differentiable. Taking advantage of that, we used the statistic defined in \citet{Contardo2022} that estimates the ``gappiness'' of each point for differentiable distributions. We constructed the statistic as follows.

For each point in the $V_R-V_\phi$ grid, we computed the gradient vector $\nabla p$ and the Hessian matrix $H$ of the density on the velocity distribution. Then we calculated the projected second derivative tensor $\Pi H \Pi \,$, with
\begin{equation}
    \Pi \equiv \mathbb{I} - \frac{\nabla p\,{\nabla p}^\top}{ {\nabla p}^\top {\nabla p} } ~,
\end{equation}
where $\mathbb{I}$ is the identity. The maximum eigenvalue of $\Pi H \Pi \,$ is the ``gappiness'' estimator of the point. Notice that ``gappiness'' is the opposite of ``overdensityness'', therefore the minimum eigenvalue of $\Pi H \Pi \,$ gives information on the position of the overdensities in the velocity distribution. For more details, we refer to the original paper \citep{Contardo2022}. 
We note that it would be ideal to apply the same methodology to the Gaia data. However, we aim to reach distant regions of the disc, where the Poisson noise is very significant and we do not have the differentiability conditions to apply this method. We tested the difference between both methods in the fiducial models, and found it to be below $1\,\kms$ in the main overdensities, well below the WT size used in the data. When the overdensities are not sharp enough horizontally, the different methods may diverge. However, these overdensities are not studied in this paper, since we focus on the main structures. For less significant structures, a careful discussion on the bias induced by the method would be needed. 

In the simulations, we applied the same methodology as in the data, described in \ref{sect:methodology}, but using the ``overdensityness'' in $V_R-V_\phi$.
This reveals a complex velocity field, rich in \emph{overdensities}, the equivalent in the simulations of the \emph{moving groups}.
To then measure the gradient of each overdensity, we first computed auxiliary velocity distributions in the radial direction (between $7$ and $9$\,kpc, in steps of $0.1$\,kpc) and the azimuthal direction (between $20$ and $40$\,deg in steps of $1$\,deg). For each BI-generated $V_R-V_\phi$ map, we detected the overdensities present and used the same methodology that we used for the data to find the set of \emph{structures}, groupings across configuration space with a single $V_R$, that compose them.

In Sect.\,\ref{sect:results_2}, we explore the parameter space of the models. For this particular case, the amount of data we generate would be too large to deal with. Instead, we fitted a parabola in $V_R-V_\phi$ for each overdentisty of interest and selected the maximum as their representative. We note, however, that sometimes one overdensity might get split into two independent ones after crossing a certain radius, for instance, through the mechanism explained in \citet{Dehnen2000} for the OLR. To avoid confusion, we select the representative obtained from the larger of the resulting overdensities after the split. Overall, this representative selection is feasible thanks to the stability of the overdensities in the simulations, which remain bounded along the configuration space, but it is infeasible for the data due to the unstable shape of the moving groups once we move a few kpc away from the SN.

%
%

\section{Data vs Fiducial Models}\label{sect:results_1}

In this section, we explore the phase-space and the spatial gradients of the kinematic structures in the SN for the \emph{Gaia} DR3 data and the models.

\subsection{Global ridges in radius and azimuth}\label{sect:global_ridges}

\begin{figure*}
    \centering
    \includegraphics[width=.95\textwidth]{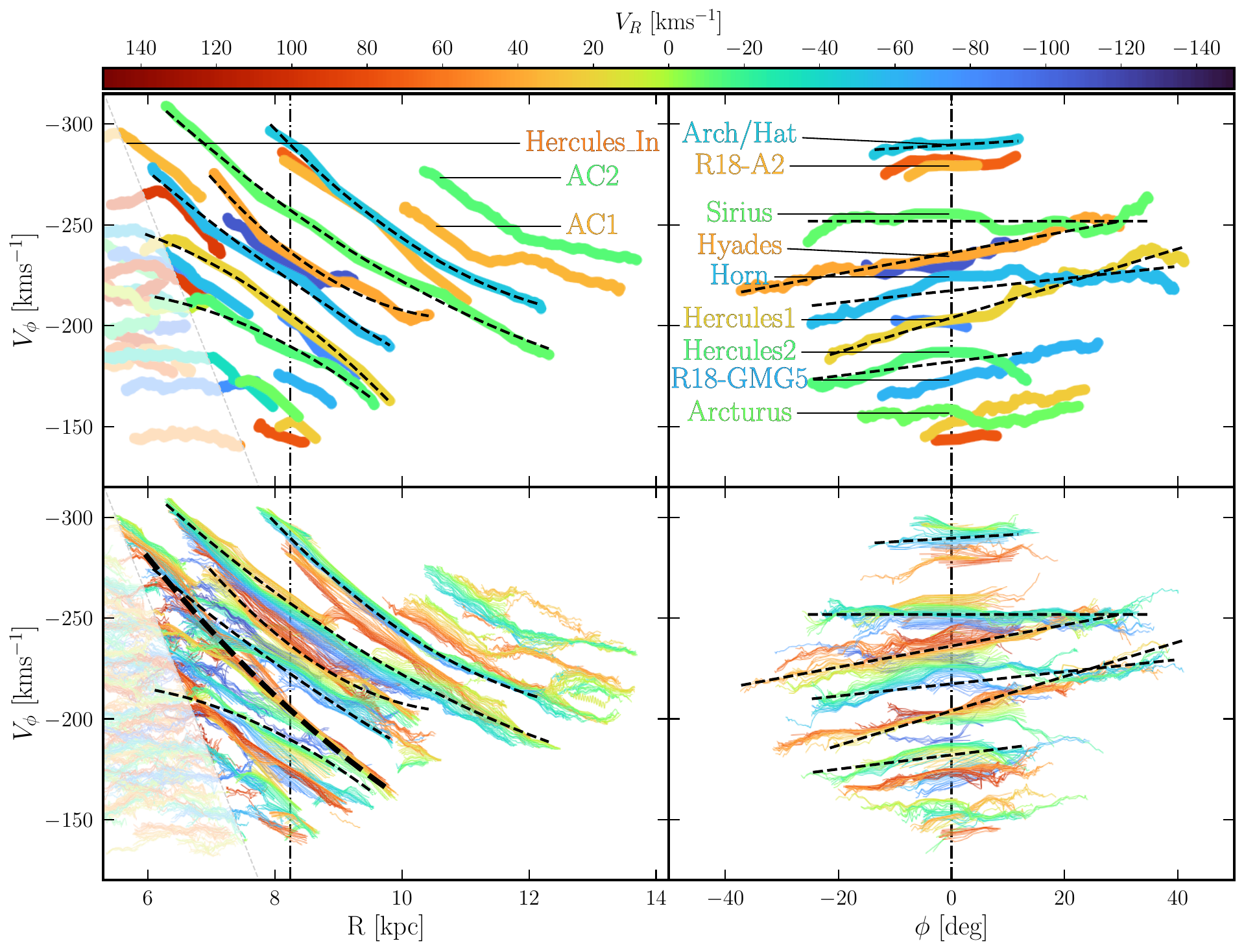}
    \caption{Azimuthal velocity of the kinematic substructures as a function of radius ($\phi = 0^\circ$, left panels) and azimuth ($R=8.24$\,kpc, right panels), and coloured by their radial velocity. The vertical error in these measurements is the size of the WT, $\sigma = 6.25\,\kms$. \textit{Top}: main structure for each moving group,  tagged with the name from the literature. The dashed black lines illustrate the fitting used to compute the gradients in the SN. \textit{Bottom}: all the structures detected by the methodology.} 
    \label{fig:vphi_r_phi}
\end{figure*}

Figure \ref{fig:vphi_r_phi} shows the structures obtained with our methodology, which trace the skeleton of the phase space. We show their $V_\phi$ in the radial ($\phi = 0$\,deg, left panels) and azimuthal ($R=8.277$\,kpc, right panels) directions. This projection allows us to study all the structures at the same time. We see how the different lines organise in bundles related to each moving group.
In the inner disc, the high kinematic temperature and the low number of stars observed blur these structures. 
Despite the usefulness of our methodology elsewhere, in this region it clearly struggles to obtain reliable detections. Thus, we ignore the detections with  $V_\phi > 80 R - 740\,\kms$ in this analysis
(translucent white region in Fig.\,\ref{fig:vphi_r_phi}, left panels).

In the first row of Fig.\,\ref{fig:vphi_r_phi}, we show the $V_\phi$ of the moving groups as a function of $R$ and $\phi$, selecting, from all the structures in each moving group at different $V_R$, only the one that covers a largest area (accounting for both the radial and azimuthal directions).
This subset selection is used only for visualization purposes. In the second row, we show all the structures that are obtained (about $1500$). 
The new results show an improvement in the resolution and extent of the structures with respect to \citetalias{Bernet2022}. Good examples of this are the extended range of detection of Hercules in the outer disc, the new structure below Sirius in the high $|V_\phi|$ part of the inner disc, and the azimuthal extension of Hyades and Sirius up to $\pm40$\,deg.

In the second row, we observe how the different bundles of lines reveal the rich ridge-like pattern. In the outer parts of the disc, we observe two clear ridges (AC1, and AC2). These were first detected in \citet{antoja2021anticentre}, where it was hypothesised that, due to the likely weaker effects of the bar resonances at these radii, they could be due to spiral arms and/or the interaction with external satellites. Below these ridges we observe the Hat, which in the radial direction, is traced robustly from the SN to $R=13.5$\,kpc.
In the azimuth direction, we are only able to detect the Hat in a small angular range. 

Sirius is the most dominant structure in the radial direction. It is detected in almost the entire studied range robustly, both in the radial and azimuthal directions.
Below Sirius, we observe Hyades and the Horn, which are especially well detected in azimuth. Hyades shows a robust linear behaviour within $\pm40$\,deg, while the Horn shows a negative slope for $\phi<0$ and a flat profile for $\phi>0$.

Finally, we focus on Hercules, which is known to be formed by three thin arches \citep{katz2018dynamics,Asano2020}. In this execution, due to the use of a WT with a larger scale, two of these arches \citep[A8 \& A9 in ][]{ramos2018} appear joined as Hercules1, while the other arch (A10 + A11) is detected and labelled as Hercules2. We note that the azimuthal velocity of Hercules1 has a clear linear increase with azimuth. As for the radial behaviour, with our methodology, both in \citetalias{Bernet2022} and here, we observe a ridge that flattens as we move to the inner parts of the disk. In \citetalias{Bernet2022}, we discussed that this flattening could be related to the influence of the centroid of the distribution in the peak detection of the methodology. Interestingly, in this work, a new set of structures have appeared in the inner disk, which we label Hercules\_In.
This structure appears as a symmetric parabola (peak of $|V_\phi|$ at $V_R = 0\,\kms$), in good agreement with Fig.~4 of \citet{Dehnen2000}, where x$_1$(2) orbits dominate (his Fig.~7).
In the $R-V_\phi$ and $V_R-V_\phi$ maps, we observe a potential continuity between Hercules1 and Hercules\_In, thus, we manually linked both sets of structures at each $V_R$ (thick dashed line in Fig.\,\ref{fig:vphi_r_phi}, bottom left panel). In the rest of the paper, we will consider these grouped structures as Hercules1. We find that, in doing so, the resulting structure is more coherent with the results obtained in the literature simulations \citep[e.g. Fig. 13 in][]{hunt2019signatures}.

\subsection{Gradients of the moving groups and overdensities}\label{sect:SN_gradients}

\begin{figure*}
    \centering
    \includegraphics[width=0.99\textwidth]{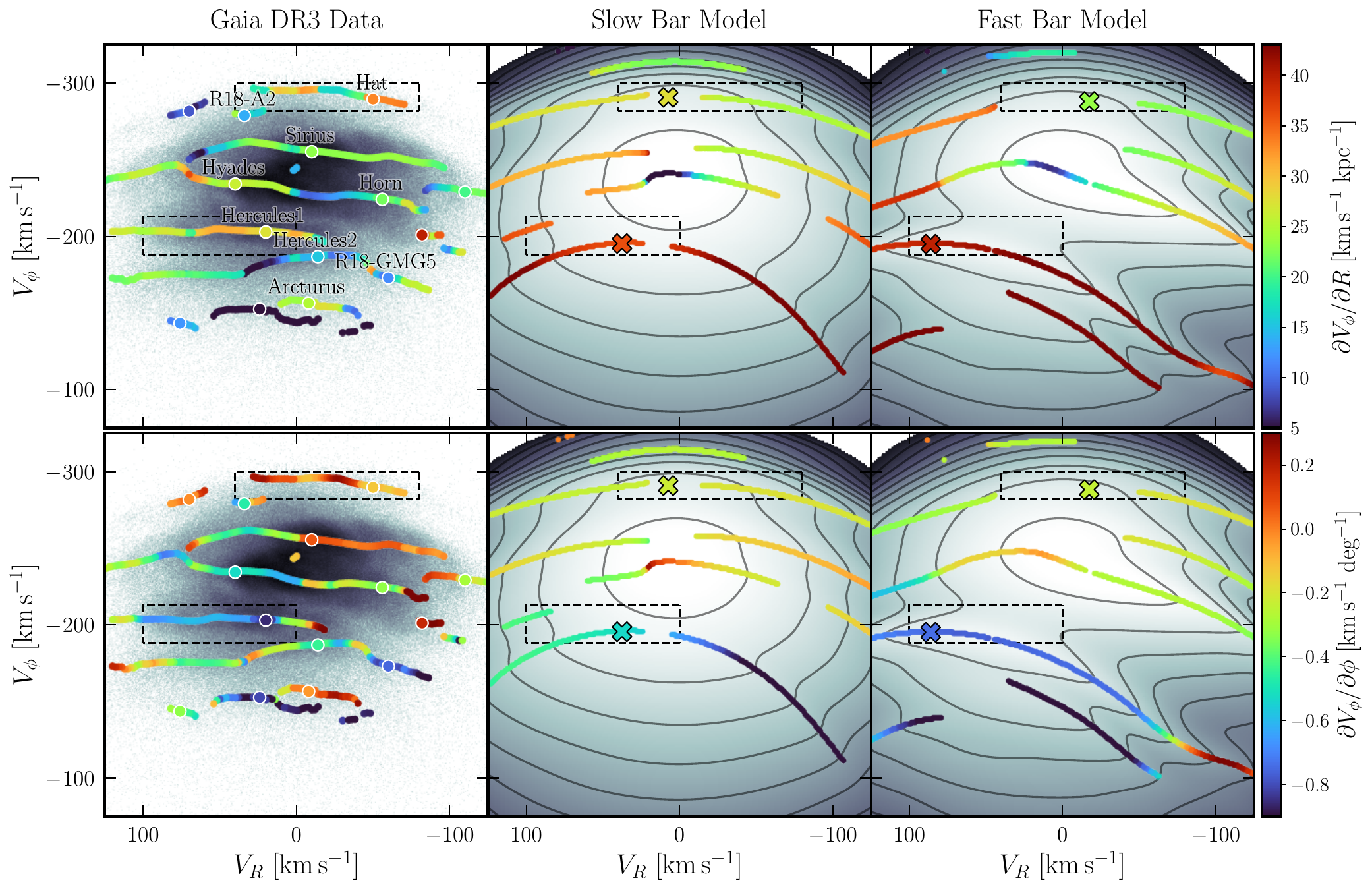}
    \caption{Moving group and overdensities detection, colored by their \dR (top panels) and \dphi (bottom panels) at each $V_R$. The dashed boxes represent the region of the velocity distribution associated with Hercules and Hat. \textit{Left}: \emph{Gaia} DR3 SN moving groups, tagged by their literature names. \textit{Middle}: SBM detection of the overdensities. \textit{Right}: FBM detection of the overdensites.} 
    \label{fig:DR3_gradients}
\end{figure*}

\begin{table}[]
\setlength{\tabcolsep}{4pt}
\caption{
Mean slope and the dispersion within the box (see Fig.~\ref{fig:DR3_gradients}) in the data and both fiducial models.}
\label{tab:slopes}
\centering
\renewcommand{\arraystretch}{1.5}
\begin{tabular}{cccc}
\hline \hline
  & & \dR{ }($\sigma$) & \dphi{ }($\sigma$)\\ 
  & & [\kmskpc] & [\kmsdeg] \\ \hline
\multirow{3}{*}{Hercules} & Data & 28.1 (2.8) & -0.63 (0.13) \\
                          & 39   & 36.5 (1.1) & -0.49 (0.05) \\
                          & 56   & 40.5 (0.9) & -0.76 (0.01) \\ \hline
\multirow{3}{*}{Arch/Hat}     & Data & 28.7 (6.8) & 0.07 (0.13)  \\
                          & 39   & 25.7 (1.4) & -0.20 (0.02) \\
                          & 56   & 22.7 (0.1) & -0.20 (0.01) \\ \hline
\end{tabular}
\renewcommand{\arraystretch}{1}
\end{table}

Figure \ref{fig:DR3_gradients} shows the moving groups detected at the SN in the \emph{Gaia} DR3 data (left panels), and the overdensities in the fiducial models (middle and right panels), colored by their gradients \dR{ }(top) and \dphi{ }(bottom). Thus, we are including information of the large-scale shape, i.e. how the velocities of the structures change with $R$ and $\phi$.
In the left panel, the big dots correspond to the largest structure of each moving group, i.e. the structures shown in the top panel of Fig.\,\ref{fig:vphi_r_phi}, which we included for visualization purposes only. In the middle and right panels, the crosses correspond to the representatives of the overdensities in the simulations, described in Sect.\,\ref{sect:BI_arch_detection}. These are the representatives used in Sect.\,\ref{sect:results_2}.

As explained in the introduction, Hercules is a moving group commonly associated with bar resonances. In the slow bar scenario, the Hat has been related with the OLR \citep[e.g.][]{monari2019signatures}. In the fast bar scenario, it is associated with the 1:1 resonance \citep{Dehnen2000}. In Fig.\,\ref{fig:DR3_gradients} we represent the regions of these moving groups with dashed black boxes. We computed the mean and standard deviation of the gradients of the structures within the boxes in the data and the models (Tab.~\ref{tab:slopes}). We do this to capture the gradients of the entire moving group, not just at a specific $V_R$, as it allows a more robust comparison with the simulations.

In Hercules1, we measured a similar azimuthal gradient throughout the entire moving group \dphi\,$\sim-0.63$\,\kmsdeg (Table \ref{tab:slopes}), compatible with both models within the error bars. In \dR, we detected two sub-segments for Hercules1: in the $V_R < 50\,\kms$ part of the arch, we measured \dR$\sim29$\,\kmskpc, while in the $V_R > 50\,\kms$ part we obtained \dR$\sim25$\,\kmskpc.
The overall average slope of Hercules1, without separating into segments, ends up being \dR$=28.1\pm2.8$\,\kmskpc. The corresponding values of \dR{ } for the SMB (36.5\,\kmskpc) and FBM (40.5\,\kmskpc), however, are both about 10\,\kmskpc larger than the gradient measured in the data.
In Section \ref{sect:results_2} we explore this discrepancy in more detail.
The gradients of Hercules2 are significantly different than Hercules1 but no counterpart is found in the models.

In the Hat, we observe different \dR{ }values within various parts of the same arch. It is important to note that the data detections in the Hat are not as reliable compared to the Hercules data, due to the low number of sources in the border of the distribution. Despite this, the gradients seen in both models are compatible with the data distribution.

The gradients in Sirius, Hyades, and Horn behave smoothly within large parts of the moving groups, with sudden breaks in some places. These breaks are a projection of the pattern already observed in Fig.\,\ref{fig:vphi_r_phi}, where on the large scale the ridges break and merge. 

We also explored the dependence of these gradients with the integration time. In Fig.\,\ref{fig:sn_t14} and Table\,\ref{tab:slopes_t14} we show the equivalent of Fig.\,\ref{fig:DR3_gradients} and Table \ref{tab:slopes} for $t_2 = 14$\,bar laps, in contrast to the $t_2 = 4$\,bar laps in the fiducial models. 
We observe that the gradients in Hercules and Hat are very similar at short and long integration times, thus indicating that time is not playing a major role in the value of these measurables, similar to \citet{Dehnen2000}, where he pointed out that the velocity of Hercules does not strongly depend on time. In the FBM, the Horn structure is maintained for long integration times, while it is not seen in the SBM.

Finally, in the negative $V_R$ part of the distribution we note a resemblance, both in position and gradients, between the overdensities in the models and Hat, Sirius, and Horn. In Section \ref{sect:azimuth_time_simil}, we study these overdensities in the models to explore an alternative hypothesis on the origin of these moving groups.

%
%

\section{Model Parameter Exploration}\label{sect:results_2}

In the literature, the resonant origin of the moving groups is usually tested by placing a given resonance (or set of resonances) in the SN, and comparing the resulting $V_R-V_\phi$ distribution with the data. We have already discussed in the introduction that different pattern speeds create resonances compatible with the data (slow-fast bar degeneracy). Other parameters can also shift the position of the overdensities in the models, thus leading to extra degeneracies, e.g. the slope of the rotation curve. In this section we use the azimuthal and radial gradients as large-scale measures and sweep different parameters of the fiducial models, to test if we can indeed break some of these degeneracies. We start by studying how the properties of each simulated structure change with the model parameters and identify properties that may strongly depend on the model (Sect.\,\ref{sect:sweep}). Secondly, we compare the properties of the models with those of the data and try to identify the parameter region that offers a better match.

Apart from the model parameters that we want to explore, there is a bunch of kinematic structures from these models to compare to the data. For this analysis we focus on Hercules and Hat as possible moving groups associated to the bar's effects. Possible links between other moving groups and the overdensities seen in the models are discussed in Sect.\,\ref{sect:azimuth_time_simil}.

\begin{figure*}
    \centering
    \includegraphics[width=\textwidth]{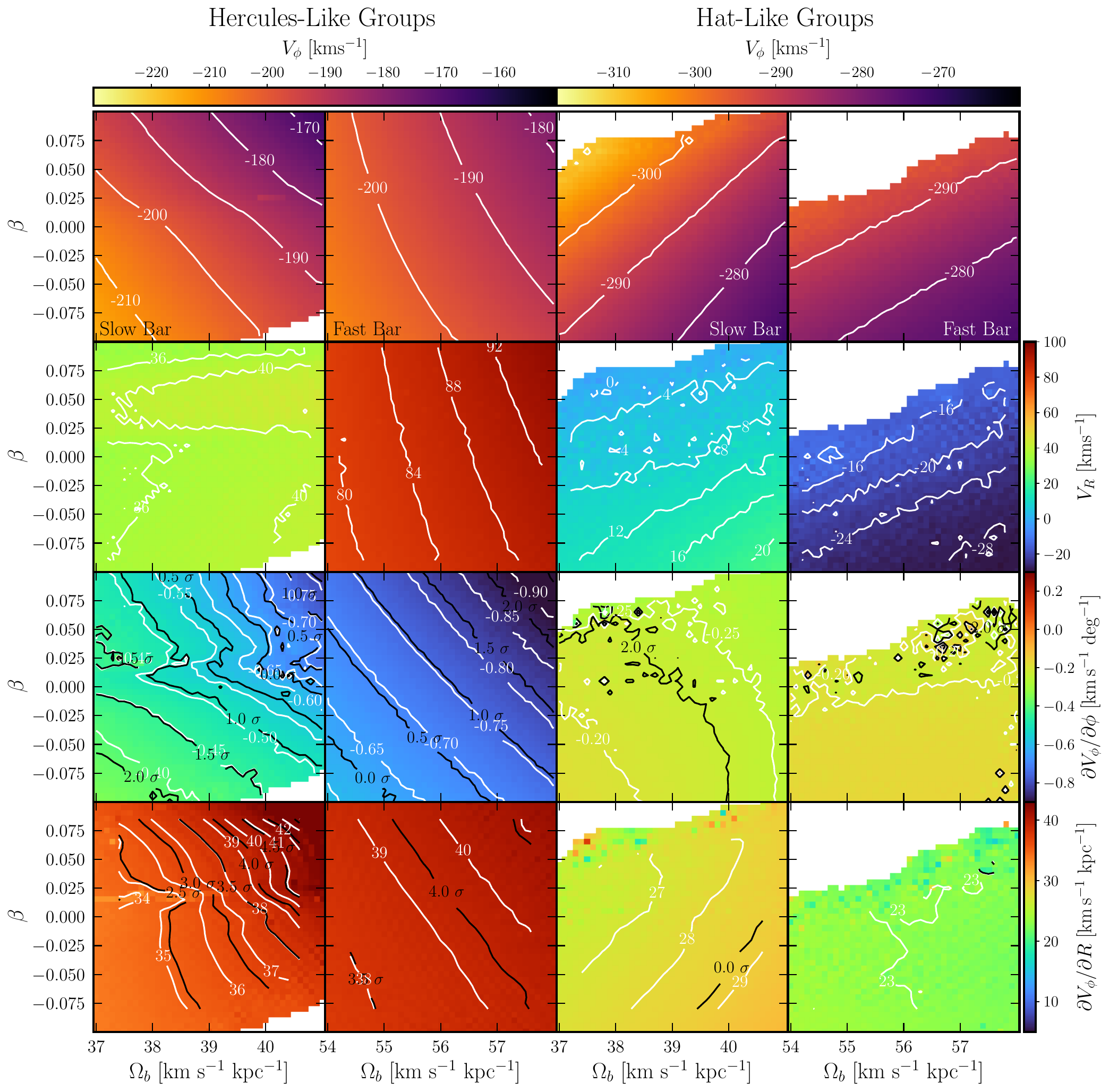} 
    \caption{Parameters of the Hercules-like (first and second columns) and Hat-like (third and fourth columns) overdensities representatives in the models. The parameter space is swept around the fiducial models $\beta$ and $\Omega_b$. \textit{First row}: $V_\phi$ of the overdensity. \textit{Second row}: $V_R$ of the overdensity. \textit{Third row}: \dphi\,of the overdensity. In black, the statistical deviation to the measurements in \emph{Gaia} DR3 (Table \ref{tab:slopes}). \textit{Fourth row}: same as third row, with \dR.}
    \label{fig:corr_p30}
\end{figure*}

\subsection{Parameter Exploration}\label{sect:sweep}

The parameter space to explore is huge and multidimensional (Table \ref{tab:simulation_params}). Below we detail each of the parameters and specify which ones we investigate:
\begin{itemize}
    \item $R$: In each model, we sweep $R$ between $7$ and $9$ kpc, in steps of $0.1$\,kpc to compute \dR.
    \item $R_\odot$: It can be fitted by external measurements \citep{Gravity2022}, so we maintain the fiducial model value $R_\odot=8$\,kpc.
    \item $v_0$: We keep this parameter fixed but we checked that changing it even by a considerable amount ($\pm15\,\kms$) does not change significantly our results (variations in the radial slopes of about 1\,\kmskpc).
    \item $\phi$: In each model, we sweep $\phi$ between $-10$ and $10$\,deg, in steps of $1$\,deg to compute \dphi.
    \item $\phi_{b}$: It can also be fitted by external measurements, though there is less consensus. In this section, we keep the value of $\phi_{b}=-30$\,deg used in the fiducial model. In Appendix \ref{sect:parameters_phi} we repeat the analysis for $\phi_{b}=-20, -40$\,deg and see that, to first order, all the trends are the same.
    \item $\alpha$ (bar strength). The strength of the bar hardly changes the position of the resonances \citep{Dehnen2000}. Therefore, we leave it fixed to the value of the fiducial models.
    \item $\Omega_p$: We explored a parameter range of $\pm 2$\,\kmskpc\,around the slow and fast bar pattern speeds, within which the $V_\phi$ of the overdensities in the models remain compatible with Hercules and Hat. 
    \item $\beta$ (rotation curve slope): We explored this parameter around the flat rotation curve ($0\pm0.1$). This range covers the realistic values for the MW rotation curve (but see discussion in Sect.~\ref{sect:discussion}).
\end{itemize}

In Fig.\,\ref{fig:corr_p30}, we show how the measurables $V_\phi$, $V_R$, \dphi, \dR{ }(different rows) of the Hercules-like (left) and Hat-like (right) overdensities behave, as a function of $\Omega_p$ (horizontal axis in all panels) and $\beta$ (vertical axis in all panels). Each pixel of each panel represents a different Galaxy Model. We cover the parameters around the two fiducial models: SBM for the first and third columns, and FBM for the second and fourth columns.

For the interpretation of some of these maps it is important to have in mind that, when a resonance moves inwards in the disc, the associated orbits reach the SN with a smaller $V_\phi$, and vice-versa. The first row of Fig.\,\ref{fig:corr_p30} shows that the azimuthal velocity $|V_\phi|$ of both the Hercules- and Hat-like groups decreases as $\Omega_b$ increases. This is expected, since the position of all resonances move inwards in the disc as the bar rotates faster. On the other hand, we see opposed behaviours with $\beta$: while the $|V_\phi|$ of Hercules-like overdensities decreases with $\beta$, that of the Hat-like overdensities increases. This can be explained by the same argument as above once we consider the fact that changes in $\beta$ affect the position of the resonances differently whether they are inside $R_0$ or outside. An increase in $\beta$ decreases the slope of the azimuthal frequency curve,
thus sending the resonances inside $R_0$ inwards, and the resonances outside $R_0$ outwards. Conversely, if $\beta$ decreases the frequencies change faster with radius, which pulls the resonances closer to $R_0$.

Regarding \dphi{ }(third row in Fig.\,\ref{fig:corr_p30}), it shows an almost linear decrease with $\Omega_b$ and $\beta$ for both Hercules-like\footnote{The Hercules-Slow Bar (leftmost column) presents an irregularity around $\beta = 0.02$ in the panels of $V_R$, \dphi, and \dR{ }due to the split in the arch (see Sect. \ref{sect:BI_arch_detection}). The global trends are maintained, so this does not affect our final conclusions.} and the Hat-like overdensities across the parameter space for the slow bar (columns 1 and 3). In the fast bar case, the gradients of the Hat-like overdensity remain at a fixed value and shows little to no variation (rightmost column).

In contrast, the \dR{} of Hercules-like structures increases almost linearly with both $\Omega_b$ and $\beta$, while in Hat-like structures, \dR{} increases with $\Omega_b$ and decreases with $\beta$. At first glance, this difference in behaviour might be related to the effect $\beta$ has on $V_\phi$, which we described above. In this sense, we can interpret all this information as follows: if the resonance moves inwards, the slope observed at the Sun increases, whether it moves inwards because of $\beta$ (increasing for inner resonances and decreasing for the outer ones) or by increasing the pattern speed of the bar. Moreover, the slope seems to be proportional to $\Omega_b$ regardless of the type of resonance or its resonant radius. It is difficult to obtain an intuition of these slopes beyond the empirical measurements, but these trends are a good milestone for an analytical modelling of these measurables. 

A visual comparison between the slow and the fast bar models reveal that, to first order, the overdensities present a similar behaviour. Specially in the Hercules-like overdensities, the contours of all the measurables are parallel, indicating a consistent directional change in the parameter space.
The only significant qualitative difference is the steepness of this change (distance between white contours) in the gradients. This similarity complicates the task of breaking degeneracies between the slow and fast bar scenarios. 

In Appendix \ref{sect:degeneracy_appendix} we study the differences between the fiducial SBM and FBM measurable for Hercules and Hat (Tab.\,\ref{tab:slopes}) and the galaxy models around the FBM. The goal is to evaluate the minimal parameters we should consider to distinguish between the models, in case the data matched the models.
We conclude that the best combination to constrain the bar parameters is to consider the measurements for both Hercules and Hat, and that including the azimuthal gradient of both structures brings the statistical significance over $8\sigma$ in almost the entire parameter space.

\subsection{Comparison with Gaia Data}

The black contours in Fig. \ref{fig:corr_p30} show the statistical deviation between each galaxy model and the data, i.e. the difference w.r.t the mean value in Table \ref{tab:slopes} in units of standard deviation. For instance, a $1\sigma$ deviation means that the difference between the model and the mean gradient of the corresponding moving group equals the standard deviation of the data reported in Table \ref{tab:slopes}.

Since we are exploring the parameter space around the fiducial models, to first order the results are similar to what we have described in Section \ref{sect:SN_gradients}. 
In some parts of the parameter space, the azimuthal gradient is matching the one seen in the data ($0\sigma$ curves). 
In the slow bar Hercules-like overdensity (leftmost column) the azimuthal gradient matches the data for $\Omega_b\sim40$\,\kmskpc and $\beta\sim0.025$. Opposite to that, the azimuthal gradient of the fast bar Hercules-like overdensity (second column) matches the data for $\Omega_b\sim54$\,\kmskpc and $\beta\sim-0.05$. 

The radial gradients, on the other hand, do not match the values of the data in any part of the parameter space explored. Thus, in general, we do not find any combination of $\Omega_b$ and $\beta$ that reproduces all the measures in the data.
To obtain a match in radial gradient we would need to decrease the parameters under $\Omega_b\ll37$\,\kmskpc, away from the match in $V_\phi$, and $\beta\ll-0.1$, well below realistic models of the MW. Similar analysis in the other azimuths (see Figs.~\ref{fig:corr_p20} and \ref{fig:corr_p40}) lead to similar differences in radial gradients, always above $3\sigma$ when comparing models and data. This confirms that our models are sub-optimal to reproduce the large-scale velocity distribution for any parameter combination.

%
%

\begin{figure*}
    \centering
    \includegraphics[width=0.98\textwidth]{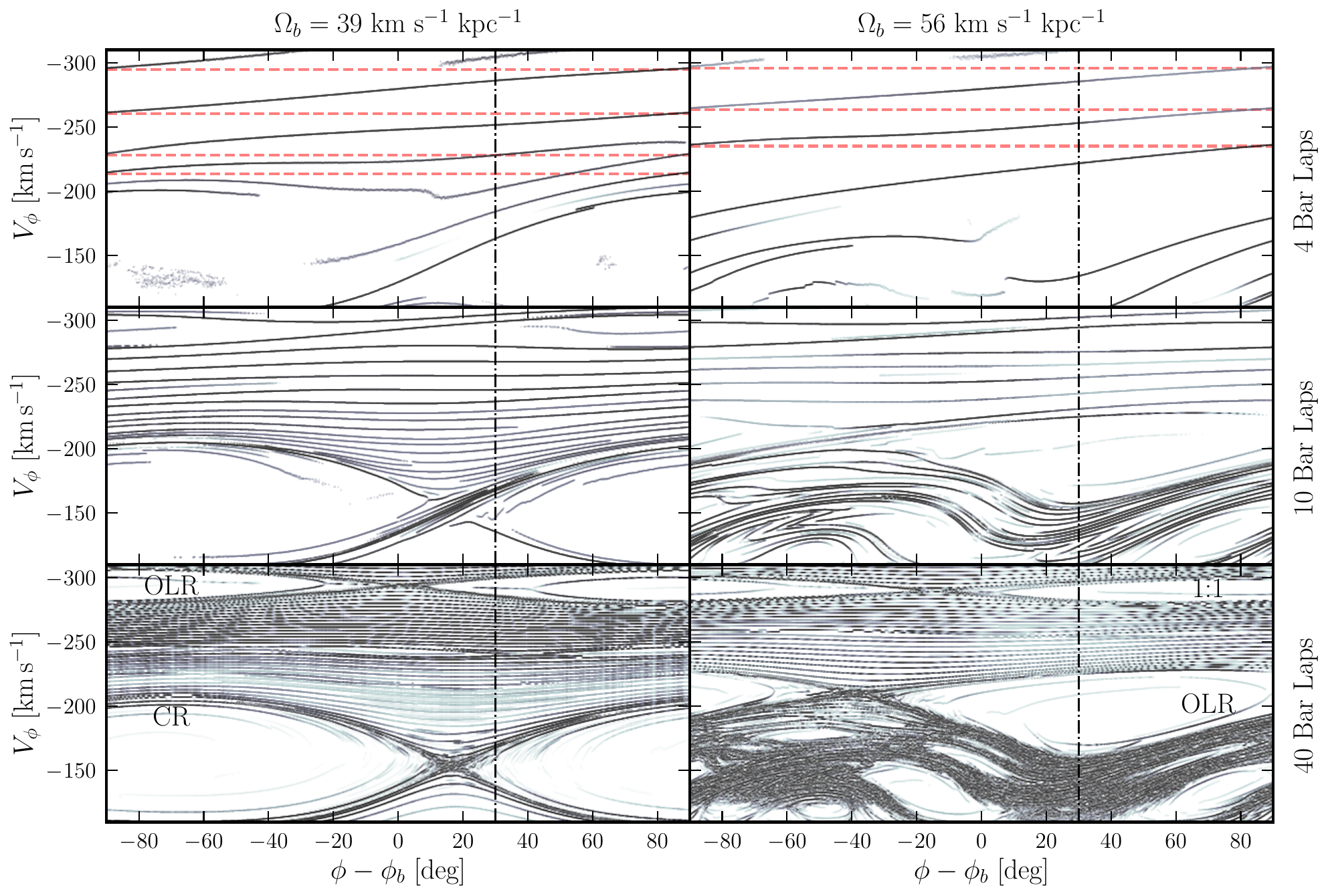}
    \caption{$V_\phi$ of the overdensities at $V_R = -60\,\kms$ for an entire turn around the model of the galaxy. \textit{Left}: Results for the SBM. \textit{Right}: Results for the FBM. \textit{Top}: Structures for $t_2=4$\,bar laps. We observe how the arches seen in the fiducial models are part of the same global structure, linked through a $180$\,deg symmetry (red dashed lines). \textit{Middle}: Structures for $t_2=10$\,bar laps. The resonant regions start to be clear, and the structures in the middle are flatter and closer between them. \textit{Bottom}: Structures for $t_2=40$\,bar laps. The resonant regions are clearly defined, and the middle region is crowded with flat structures. The vertical dashed lines mark the location of the Sun.}
    \label{fig:omega_time_phi}
\end{figure*}

\section{Azimuth and Time Exploration}\label{sect:azimuth_time_simil}

In Section \ref{sect:SN_gradients}, we pointed out the resemblance, both in position and gradients, between the overdensities in the models and the moving groups Hat, Sirius, and Horn in the data. The goal of this section is to understand these structures in the models and the potential implication on the origin of the mentioned moving groups.

\subsection{Origin of the overdensities in the models}

We aim to study the global shape of these overdensities in the negative $V_R$ region of our models, which is the region of the Horn that at the same time includes Sirius and Hat.
To simplify our analysis, we focus on a subset of structures, arbitrarily those with a radial velocity $V_R=-60\,\kms$. In Fig. \ref{fig:omega_time_phi}, we present the azimuthal velocity ($V_\phi$) of these structures across the azimuthal range of $\pm90$ degrees with respect to the long axis of the bar. The models  have $180$-deg symmetry, so we just need to cover half of the disc.
In the top panels, we study the structures in the fiducial models. We see that all the overdensities in the SN (black vertical line) belong to a few global structures, that link the different overdensities in the SN in an entire turn around the galaxy (forming double helical shapes). The dashed red lines in the top panels of Fig.\,\ref{fig:omega_time_phi} represent these links \footnote{Notice that we are assuming a $180$-deg symmetry. Therefore we are linking each structure with its counterpart.}. These connections could already be suspected in Fig.\,2 of \citet[]{Dehnen2000}, but to our knowledge, it is the first time that it is characterised in detail.

Since this analysis is done at very short timescale ($4$ bar laps, i.e. $0.64$\,Gyr for the slow bar, and $0.44$\,Gyr for the fast bar) in the second and third rows of Fig.\,\ref{fig:omega_time_phi} we do the same analysis for $t_2 = 10$ and $40$\,bar laps, respectively. We observe two phenomena: the predicted resonant regions appear (CR and OLR for the slow bar and OLR and 1:1 for the fast bar), and the global helix winds, flattening in $\phi$, except at the resonance boundaries, where the lines get squeezed.
To understand the evolution from one row to another, in Fig.\,\ref{fig:time_dR_dphi} we show the velocity of these structures with time. We observe that the $V_\phi$ of most of the structures evolves in time, asymptotically approaching the boundaries of one of the resonant regions.

In the last row of Fig.~\ref{fig:omega_time_phi} we can clearly see the nodes of the resonant regions, which are not aligned with the bar major axis as one would expect. We computed the position of the nodes for positive $V_R=50\,\kms$ and they are located in the opposite site of the major axis. Therefore, at $V_R=0\,\kms$ the nodes will be aligned with the bar as expected, indicating that the displacement of these nodes is probably due to the eccentricity of the orbits. A more detailed description of these nodes is beyond the scope of this work.

In summary, in the models we have two types of overdensities: resonances, and transient arches. The position of the resonance boundaries stabilize relatively quickly ($\sim8$\,bar laps) and remain fixed, as previously anticipated in \citet{Dehnen2000}. The transient arches are joined trough an entire turn around the disc forming a global structure with helical shape. In addition, this helical shapes ends at the resonance boundaries, joining all the overdensities. These transient arches are the result of introducing the bar potential in an axisymmetric DF, and thus contain temporal information of the model.

\subsection{Comparison with Gaia Data}

\begin{figure}
    \centering
    \includegraphics[width=0.49\textwidth]{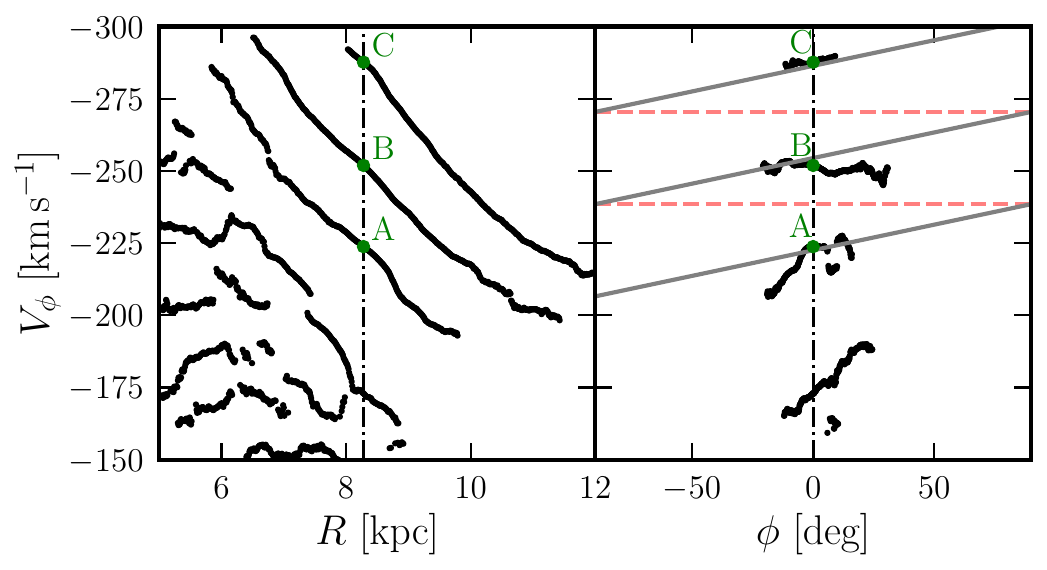}
    \caption{Detection of structures in the data at $V_R = -60\,\kms$ (extracted from the bottom panel of Fig.~\ref{fig:vphi_r_phi}). \textit{Left:} Radial position of the structures. \textit{Right:} Azimuthal position of the structures.}
    \label{fig:R-phi-Vphi}
\end{figure}

As an exercise, we now assume that, in the MW, the Hat, Horn, and Sirius moving groups are indeed related with the same global helical structure. In this scenario, the Hat would be formed in a resonance boundary and Sirius and Horn would in turn be the imprints of the ongoing phase-mixing. In Fig.\,\ref{fig:R-phi-Vphi}, we show the detection of structures in the data for $V_R = -60\,\kms$ in the radial and azimuthal direction. Following the assumption that they are related, we compute the fitting of a linear helix in azimuth (grey line on the right panel). We obtain a global slope for this structure of \dphi$=-0.177$\,\kmsdeg. Although the fitting results are poor, especially for Sirius (B), it does match the slope of Hat (A) and the parts of Horn (C) at $\phi>0$.

\begin{figure}
    \centering
    \includegraphics[width=0.49\textwidth]{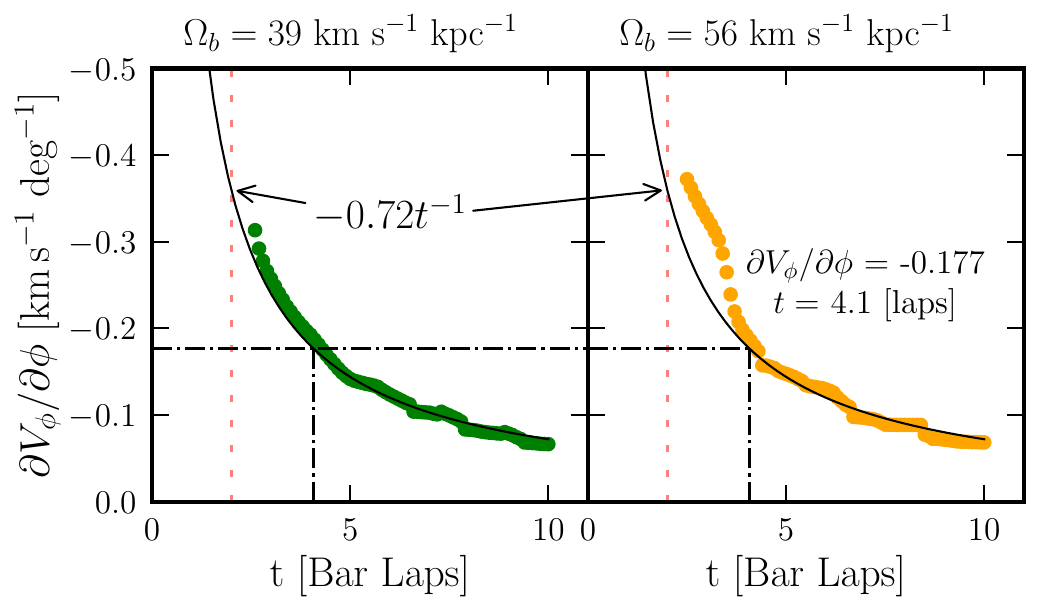}
    \caption{Azimuthal slope of the transient arches at each integration time, for a Slow and a Fast bar. The black line shows the curve $-0.72t^{-1}$, which approximates both curves. The dashed black line corresponds to the slope fitted in the data (Fig.\,\ref{fig:R-phi-Vphi}), and the intersection returns the time estimation. The dashed red line marks the end of the growth of the bar ($t=2$\,bar laps).}
    \label{fig:time_estimation}
\end{figure}

Back to the models, we observed that the space between the transient arches decreases with time (Fig.\,\ref{fig:omega_time_phi}). That is, the \dphi{ }of the transient arches in azimuth decreases with time. We note that the diagrams of this figure show aspects similar to typical phase-mixing in a frequency-angle plot (in our case equivalent to $\Omega_\phi$-$\phi$ plane), where the lines get closer and have smaller slopes with time \citep[e.g.][for the case of the phase spiral]{li2021phasespiral,frankel2023,DarraghFord2023}. In Fig.\,\ref{fig:time_estimation} we compute this slope of the transient arches for the slow (green points) and fast (orange points) bar model at different integration times. We see that, if $t$ is expressed in bar laps, the evolution of the slope with time is the same for both pattern speeds. It seems therefore that in this case the phase-mixing times are modulated not only by the orbital frequencies but also by the bar as well.

Again, assuming that the structures in the model are the same as the moving groups, we see that the measured slope for the data (see above) gives a timescale estimate of the MW bar of $t\approx 0.6$\,Gyr. This is evidently a very short timescale compared with the usual MW bar age estimations  \citep[$\gtrsim8$\,Gyr, e.g.][]{grady2020barage,sanders2022barage}. In the following section, we discuss the implications of this short timescale prediction.

%
%

\section{Discussion}\label{sect:discussion}

\paragraph{Radial gradients.}


The main disagreement between the models and the data is observed in the radial gradients of the moving groups. 
Our measurement of the radial gradient of Hercules \dR$=28.1\pm2.8$\,\kmskpc is compatible with recent results such as the one in \citet{ramos2018}, of \dR$=26.5$\,\kmskpc, but does not match the ones in the models by a significant amount (Table \ref{tab:slopes}).
In previous studies, they found a good agreement between the data and simulations in the case of the FBM. For instance, in \citet{antoja2014constraints} the detection of Hercules\footnote{Technically, the gap between Hercules and Hyades-Pleiades.} in the RAVE data at different radius matched the empirical predictions of test particle models. Revisiting this work, we note that the radial slope measured with RAVE (\dR$=\sim33\pm10$\,\kmskpc) was significantly larger than the one found in this work but with a large measurement error. In our case, even when modifying the slope of the rotation curve (row 4 in Fig.\,\ref{fig:corr_p30}) or the angle of the bar (row 4 in Figs.\,\ref{fig:corr_p20} and \ref{fig:corr_p40}) we are still far from obtaining a match between the models and the data, with a disagreement above $3\sigma$. However, it could be possible that the MW circular velocity curve is not well approximated by a single power-law. In fact, as discussed in \citet{antoja2022wave} where we examined various models and observations for the MW, not all cases could be well fitted by a single power law model. For instance, the model by \citet{McMillan2017} shows raising and decreasing trends with the transition at $R\approx 8\,\kpc$. Quantifying this effect with the current set-up requires changing some of its pieces, like the DF used. However, we can use the formalism of Ramos et al. (in prep.) to assess the impact of different rotation curves on the slopes.

It may be that a more sophisticated axisymmetric model, the effect of spiral arms \citep{hunt2018transient}, self-gravity, Giant Molecular Clouds (GMCs), external perturbations, and/or, the scenario that is currently gaining the most traction, a decelerating bar \citep{chiba2021decelerating,chiba2021decelerating_ring} are non-negligible effects that need to be taken into account. A clear next step using the framework that we have constructed is to compute the radial gradients of the moving groups in a decelerating bar and check if we are able to reproduce the trends measured in \emph{Gaia}.

\paragraph{Azimuthal gradient of Hercules and Horn.}

In this analysis, we measure an azimuthal slope for Hercules of \dphi$=-0.63\pm0.13$\,\kmsdeg. Using also \emph{Gaia} DR3\footnote{They use \citet{bailerjones2021} distances and a $R_0=8.15$\,kpc instead of $R_0=8.277$\,kpc used here. This can modify the value of the measurement, but the global trends are not expected to change significantly.}, in \citet{lucchini2023b} they obtain a measurement of $-0.74\pm0.04$\,\kmsdeg. In this work, the $V_R$ is marginalized, obtaining a smaller error than ours, but loosing the $V_R$ information. Similarly, in \citetalias{Bernet2022} we computed \dphi$=-0.5$\,\kmsdeg using \emph{Gaia} EDR3 data.

Using a combination of perturbation theory and the pendulum approximation \citep{monari2019signatures}, in \citet{monari2019hercules} they found a slope of $\partial L_z / \partial \phi = -8$\,kms$^{-1}$kpc\,deg$^{-1}$ at a radius of $8.2$\,kpc, which corresponds to \dphi$=-0.96$\,\kmsdeg, when looking at the mean $V_R$ of the model of a slow bar, which roughly agrees with the observations. They also observe the Horn variation with azimuth, which we confirmed in Figs.\,\ref{fig:vphi_r_phi} and \ref{fig:DR3_gradients}. In \citet{monari2019hercules}, they claim that in the fast bar regime, a Horn moving group with an OLR origin would show a flat azimuthal structure. In contrast, in the slow bar regime, a Horn created by the 6:1 resonance would have a significant slope. In the case of Hercules, they predict a distinct azimuthal slope in the slow bar regime (CR origin) and a less pronounced slope in the fast bar regime (OLR origin).

Our measurements in the fiducial models show evidence that both the SBM and the FBM present a significant gradient in azimuth for the Hercules-like overdensity (Table \ref{tab:slopes}) for short and long integration times. We also confirm that the azimuthal structure of the fast bar Horn-like structure is flatter than the Hercules-like. However, due to the absence of high-order moments in our simple bar model, we can not generate a Horn-like overdensity (6:1 resonance) in the slow bar regime.

We observe that for Hercules-like overdensities, both azimuthal gradients are still within the error bar of the data (even when exploring the parameter space around the fiducial models, see row 3 in Fig.\,\ref{fig:corr_p30}). 
We would need to reduce the observational errors at least by a half to obtain a significant similitude (or difference) between the models and the data (Tab.\,\ref{tab:slopes}). Finally, we note that the measure in \citet{lucchini2023b}, with a smaller error, is very compatible with our models with a fast bar.

\paragraph{Azimuthal shape of Hercules.}

In an ideal, long-lived barred galaxy, the resonances form steady state closed islands in the $V_\phi-\phi$ diagrams (Fig.\,\ref{fig:omega_time_phi}) with a certain azimuthal periodicity, depending on the resonance order \citep[e.g.][]{tsoutsis2009,michtchenko2018resonances}. The features in the velocity distribution related to these islands will then follow periodic curves in azimuth. Therefore, either they have null slope or they must contain a minimum and a maximum within a period \citep[Bolzano's theorem,][]{bolzano1817}. If Hercules is related to a resonance, it should follow one of these periodic curves in the $V_\phi-\phi$ space. Thus, it should contain a maximum and a minimum in $-90\leq\phi\leq90$\,deg.

In our measurements, we observe Hercules to follow a linear trend covering up to $60$\,deg and we do not observe a maximum nor a minimum. There are two possibilities: the maximum and the minimum are located in a part of the disc that we have not observed yet, or the resonance regions are not stationary (closed), indicating that the MW disc is far from phase-mixed.
Future spectroscopic surveys (WEAVE, 4MOST, SDSS-V/MWM) covering larger fractions of the disc will allow us to detect Hercules in a larger azimuthal range, ideally going beyond the long axis of the bar, and confirm its connection to this non-axisymmetric component of the Galaxy.

\paragraph{Origin of Hat, Sirius and Horn.}

In Sect.~\ref{sect:azimuth_time_simil}, we used the moving groups Hat, Sirius and Horn (in particular, their azimuthal gradient) to time the stage of phase-mixing induced by the growth of the bar. Our estimate of $\sim0.6$\,Gyr contradicts by far the current measurement of the age of the MW bar, $\gtrsim8$\,Gyr \citep{grady2020barage,sanders2022barage}. Thus, given the initial assumption, it is unlikely that our hypothesis is true. However, these timescales below one Gyr would be compatible with more recent events, such as the Sagittarius (Sgr) impact, and a paradigm of ongoing phase-mixing of the MW disc. In fact, other studies have determined the phase-mixing timescales in the planar velocities ($V_R$-$V_\phi$) following a perturbation from a galactic satellite. These studies find similar but slightly larger times. For instance, in \citet{antoja2022wave} we found times of $0.8-2.1$\,Gyr based on the separation between peaks of the wave in $V_R$-$L_Z$. Prior to that, \citet{minchev2009milky} found a start of phase-mixing time of $2$\,Gyr based on a separation between moving groups of $20\,\kms$. For this same velocity separation, from our Fig.\ref{fig:time_estimation} we would obtain times of around 1 Gyr, i.e. smaller than in \citet{minchev2009milky}, and which also depend on the pattern speed of the bar, thus indicating a different underlying phase-mixing mechanism. Although the studies mentioned are more specific for an external perturbation than the present work, they do not explore the azimuthal dimension and are still simplified models. More tests must be done to establish the relation between certain moving groups and Sagittarius, both with idealized models \citep{antoja2022wave} and taking into account the effect of self-gravity \citep{Darling2019}.

\paragraph{Possible caveats of this work}

The main contribution of this work is the development of novel techniques to quantify the large-scale kinematic substructure in both the data and models. The final goal of this quantification is to perform a direct discrimination between the  likelihood of the models. However, this task has a clear obstacle: a proper modelling of the data. The models used in this work, despite being intentionally simple, depend on a large set of parameters. Throughout the present work, we tested the impact on the measurables of each of the parameters of the model (Tab.\,\ref{tab:simulation_params}) and the ‘observational’ methodology (vertical size of the selection, overdensity detection, gradient fitting…). We assessed that the bias produced by each of the tested factors independently is well below the precision we can measure in the actual data. However, there is a non-negligible possibility that a combination of biases in the model and/or the methodology is the cause of the clear disagreement with the data. Regardless, as mentioned above, the simplicity of the model itself is most likely the main contributing factor, as the interaction with a satellite, spiral arms, and/or a slowing bar would play a major role that has not been tested in this work.

%
%

\section{Summary and conclusions}\label{sect:conclusions}

In this work we provide a detailed analysis of the kinematic substructure of the MW disc in the \emph{Gaia} DR3 data across $7$\,kpc in radius and $80$\,deg in azimuth. We compared the results in the data
to a suite of BI simulations, and we found that the overdensities detected in the fiducial models, as well as their gradients to a lesser degree, are compatible between them, and that  this similarity is maintained even in the parameter space around the fiducial models. Our main findings and conclusions are:

\begin{itemize}
    \item We detect new ridges in the outer disc, propose a new radial profile for Hercules, and precisely characterise Hyades, Horn, and Hercules $\sim70$\,deg along the disc.
    \item We observe a robust slope in azimuth for Hyades and Horn, previously unexplored.
    \item The radial gradient of Hercules is \dR$=28\pm2.3$\,\kmskpc, and the azimuthal gradient is \dphi$=-0.63\pm0.13$\,\kmsdeg.
    \item The radial gradient of Hercules in the BI models disagrees with the data with a significance above $3\sigma$ for all the explored parameters ($\Omega_b$, $\beta$, and $\phi_b$). We conclude that more complex models, which take into account spiral arms, self-gravity, external perturbations, and/or a slowing-down bar, are required to reproduce the observations. Despite this, we acknowledge that this problem has a high complexity and we could be underestimating some biases in the analysis.
    \item The azimuthal gradient of Hercules in the data is compatible with both the slow and the fast bar models. For these simple models, we require at least half of the error in the data to disentangle between them.
    \item Our analysis points out that a robust determination of the pattern speed using our type of analysis requires fitting the azimuthal velocity and the azimuthal gradient of at least two moving groups.
    \item We study the azimuthal slope of the phase-mixing structures induced by the growth of the bar in the models, and asses that for identical Galaxy models, the evolution of the slope only depends on the pattern speed of the bar.
    \item We explore the possibility of a phase-mixing origin for Sirius and Horn in the context of the growth of the bar, which leads to a too low estimate of the bar age ($\sim0.6$\,Gyr). We conclude that a different hypothesis is required to explain these structures.
\end{itemize}

With this work we confirm the potential of our methodology and of the \emph{Gaia} data to provide a quantitative description of the MW kinematic substructure.
In the near future, the use of spectroscopic surveys to extend the range of the \emph{Gaia} 6D sample, specially towards the tip of the MW bar, will allow us to disentangle the resonant origin of Hercules. Finally, this project makes it clear that simple models can not capture the full complexity of our Galaxy, specially in the large-scale. It emphasizes the need of exploring more advanced models, and reminds that an optimal model will have to go beyond matching local patterns.

\begin{acknowledgements}
We thank the referee, Ralph Sch\"onrich, for his helpful and detailed comments, and the notably instructive and pleasant discussion with him.
This work has made use of data from the European Space Agency (ESA) mission {\it Gaia} (\url{https://www.cosmos.esa.int/gaia}), processed by the {\it Gaia} Data Processing and Analysis Consortium (DPAC, \url{https://www.cosmos.esa.int/web/gaia/dpac/consortium}). Funding for the DPAC has been provided by national institutions, in particular the institutions participating in the {\it Gaia} Multilateral Agreement. 
This work was (partially) supported by the Spanish MICIN/AEI/10.13039/501100011033 and by "ERDF A way of making Europe" by the “European Union” and the European Union «Next Generation EU»/PRTR,  through grants PID2021-125451NA-I00 and CNS2022-135232, and the Institute of Cosmos Sciences University of Barcelona (ICCUB, Unidad de Excelencia ’Mar{\'\i}a de Maeztu’) through grant CEX2019-000918-M.
This work was partially supported by the OCRE awarded project Galactic Research in Cloud Services (Galactic RainCloudS). OCRE receives funding from the European Union’s Horizon 2020 research and innovation programme under grant agreement no. 824079.
MB acknowledges funding from the University of Barcelona’s official doctoral program for the development of a R+D+i project under the PREDOCS-UB grant.
TA acknowledges the grant RYC2018-025968-I funded by MCIN/AEI/10.13039/501100011033 and by ``ESF Investing in your future''. 
BF and GM acknowledge funding from the Agence Nationale de la Recherche (ANR project ANR-18-CE31-0006 and ANR-19-CE31- 0017) and from the European Research Council (ERC) under the European Union’s Horizon 2020 research and innovation programme (grant agreement No. 834148).
\end{acknowledgements}


\bibliographystyle{aa}
\bibliography{MyBib}

\newpage

\begin{appendix}

\section{Time dependence of the arches}

\begin{figure*}
    \centering
    \includegraphics[width=0.8\textwidth]{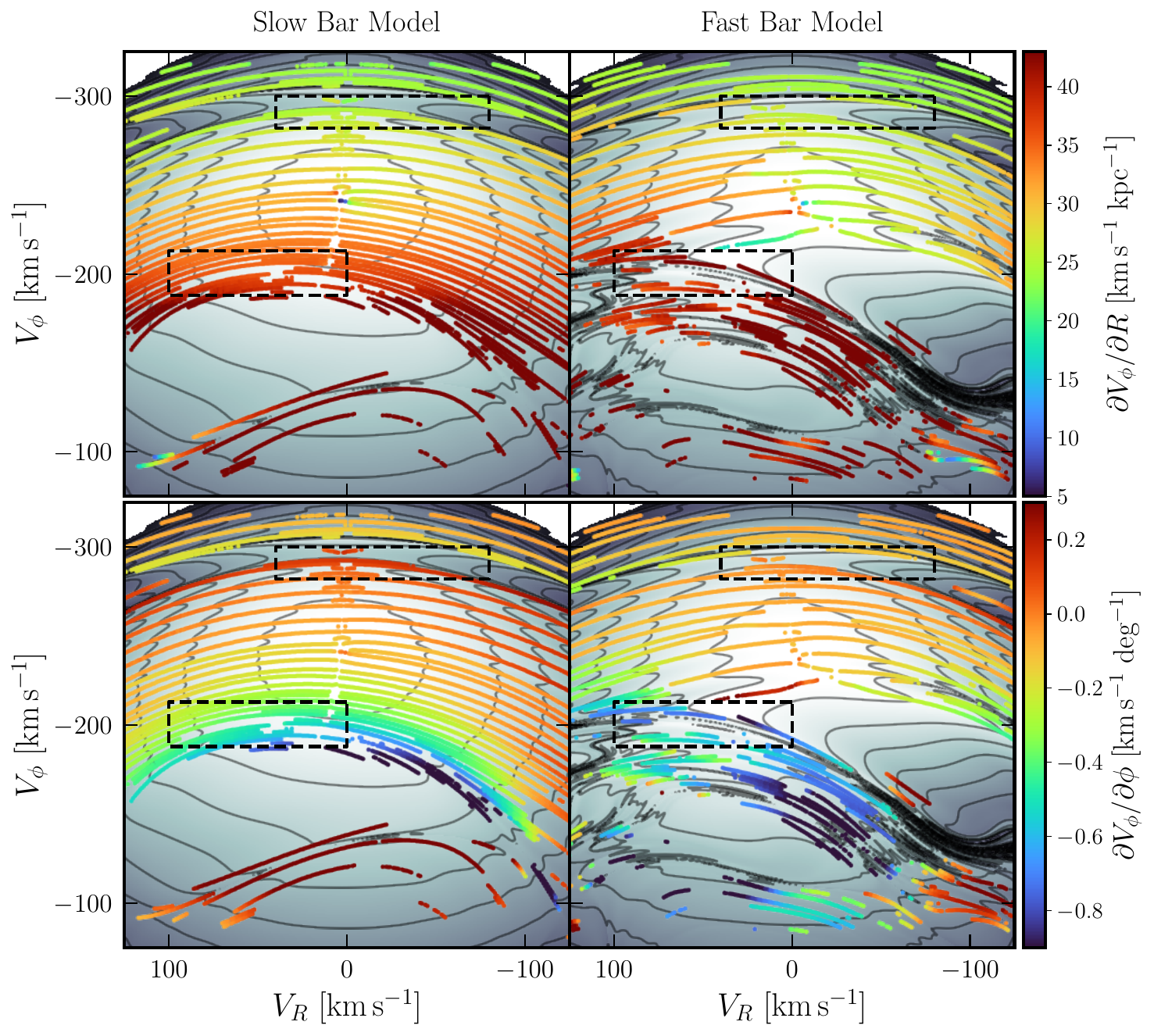}
    \caption{Overdensities detection in the simulations, for $t=14$ bar laps. Colored by their \dR (top panels) and \dphi (bottom panels) at each $V_R$. The dashed black boxes represent the region of the velocity distribution associated with Hercules and Hat. \textit{Left}: SBM detection of the overdensities. \textit{Right}: FBM detection of the overdensites.}
    \label{fig:sn_t14}
\end{figure*}

\begin{table}[]
\setlength{\tabcolsep}{4pt}
\caption{Mean slope and the dispersion within the box (see Fig.~\ref{fig:DR3_gradients}) in both models for $t = 14$ bar laps.}
\label{tab:slopes_t14}
\centering
\renewcommand{\arraystretch}{1.5}
\begin{tabular}{cccc}
\hline \hline
 &  & $\partial V_\phi / \partial R \quad (\sigma)$ & $\partial V_\phi / \partial \phi \quad  (\sigma)$ \\ \hline
\multirow{2}{*}{Hercules} & 39   & 36.2 (1.8) & -0.43 (0.11) \\
                          & 56   & 40.5 (5.7) & -0.62 (0.13) \\ \hline
\multirow{2}{*}{Arch/Hat} & 39   & 24.5 (0.9) & -0.13 (0.06) \\
                          & 56   & 24.6 (1.1) & -0.10 (0.03) \\ \hline
\end{tabular}
\renewcommand{\arraystretch}{1}
\end{table}

In order to understand how does the velocity distribution evolve in time, we integrate it $14$ bar laps and compute the radial and azimuthal gradient of each overdensity. We show these results in Fig.~\ref{fig:sn_t14} (equivalent of Fig.\ref{fig:DR3_gradients}) and Table~\ref{tab:slopes_t14} (equivalent of Tab.~\ref{tab:slopes}). We see that the both Hercules-like and Hat-like structures are located in the same position at all times. The gradients of the main structures are also very similar as the fiducial ones.

As for the presence of overdensities, the only features that are maintained are the Hercules-like and Hat-like for both pattern speeds, and the Horn-like for the fast bar. In the slow bar regime, a Horn-like structure can be produced by the 6:1 resonance, but our bar is a simple quadrupole that has no high order moments. Therefore, it is expected that these high order resonances are not appearing.

\begin{figure*}
    \centering
    \includegraphics[width=0.8\textwidth]{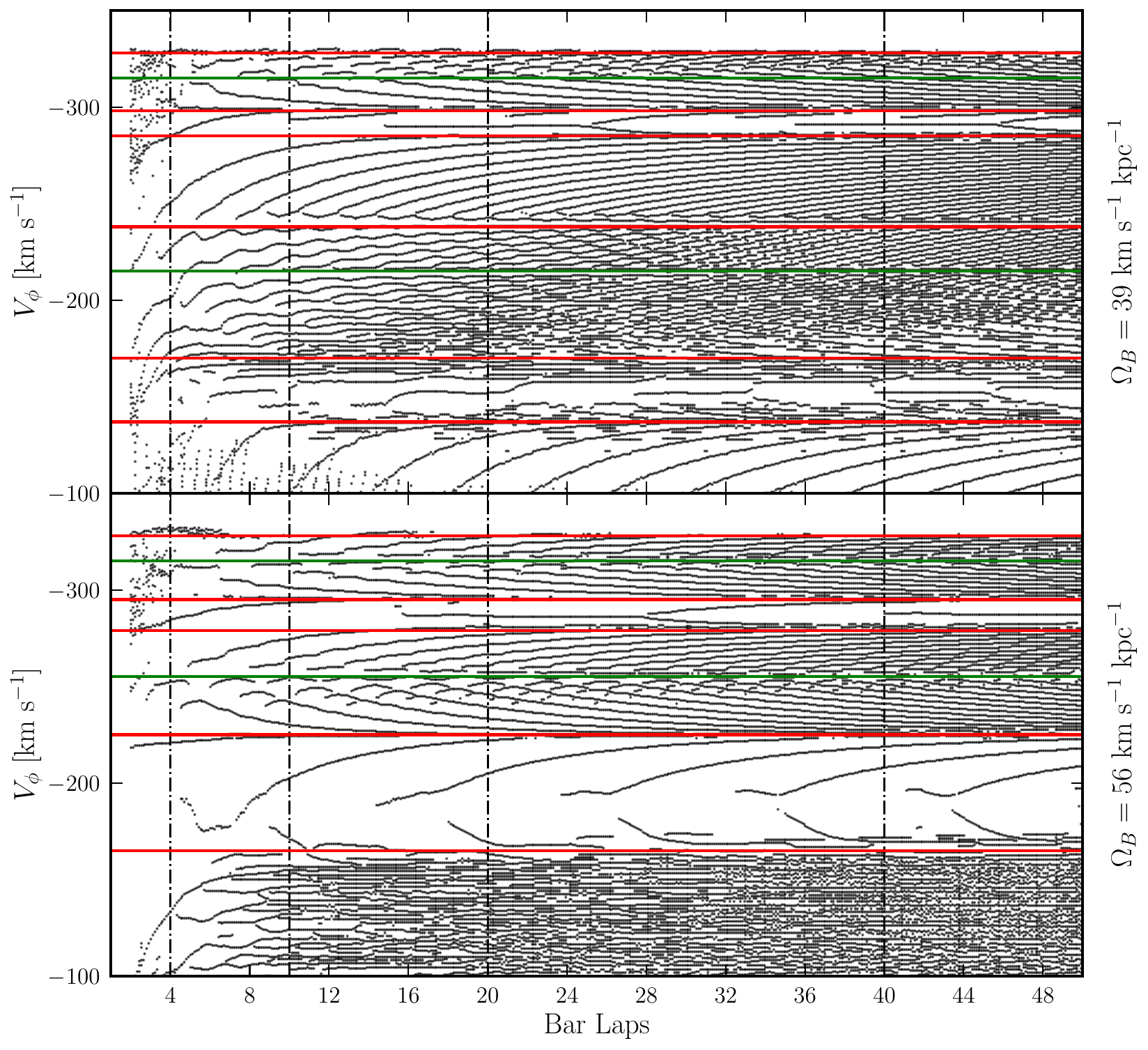}
    \caption{Time evolution of the structures at $V_R=-60\,\kms$. The structures are created in given positions of the phase space (green lines) and tend asymptotically to the boundaries of a resonance (red lines), following exponential decays in time. \textit{Top}: Slow bar model. \textit{Bottom}: Fast bar model.}
    \label{fig:time_dR_dphi}
\end{figure*}

To gain an intuition in how do the overdensities evolve from one integration time to the other, we selected the structures in $V_R=-60\,\kms$ and traced them along $50$ bar laps (Fig.\,\ref{fig:time_dR_dphi}). We observe that the structures are created in given positions of the phase space (green lines) and tend asymptotically to the boundaries of a resonance (red lines), following exponential decays in time. Understanding this behaviour could bring new intuitions to understanding the phase-mixing of this system.

\section{Correlation in other azimuths}\label{sect:parameters_phi}

\begin{figure*}
    \centering
    \includegraphics[width=\textwidth]{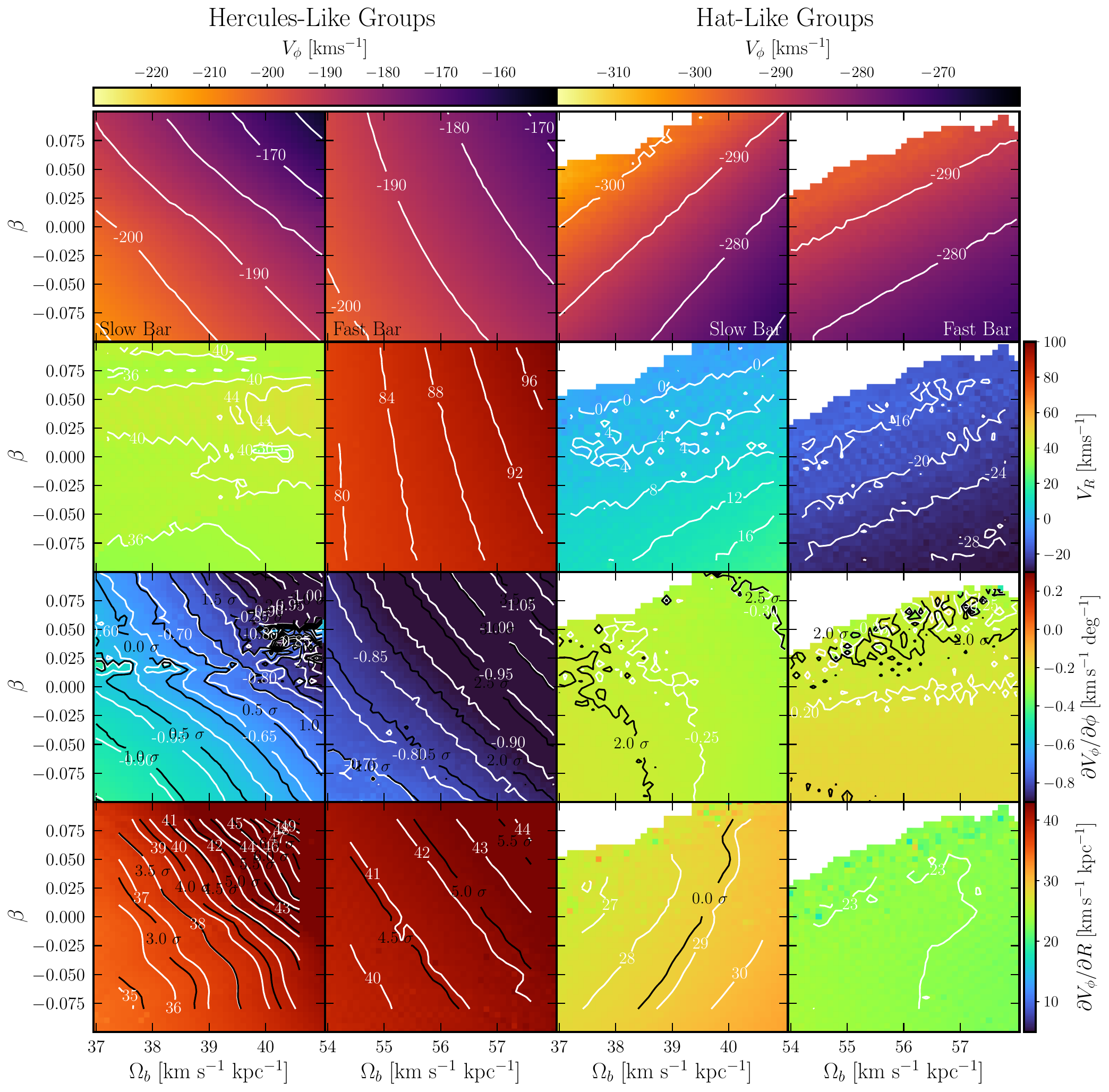}
    \caption{Same as Fig. \ref{fig:corr_p30} computed with $\phi_b = -20$~deg.}
    \label{fig:corr_p20}
\end{figure*}

\begin{figure*}
    \centering
    \includegraphics[width=\textwidth]{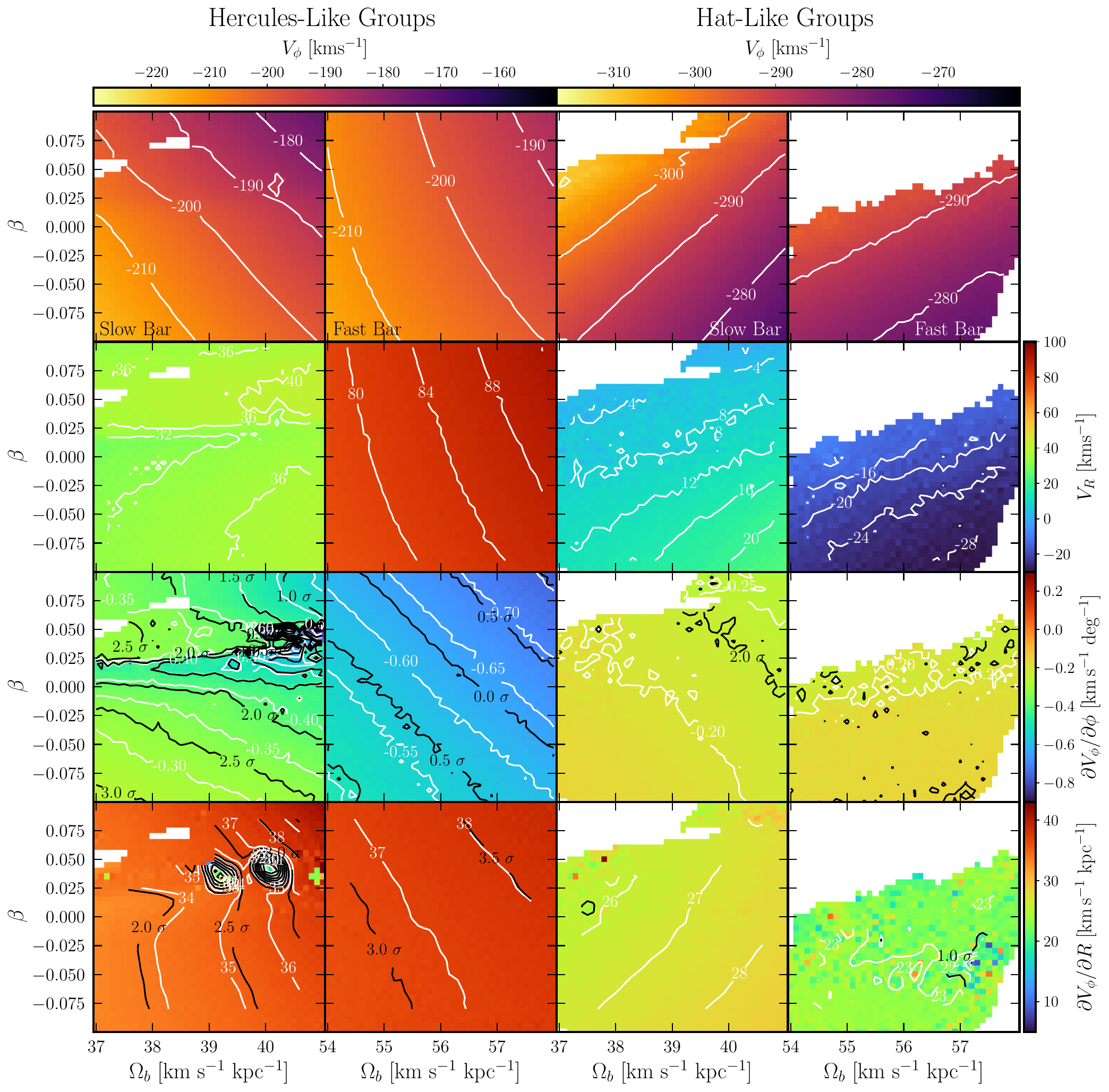}
    \caption{Same as Fig. \ref{fig:corr_p30} computed  with $\phi_b = -40$~deg.}
    \label{fig:corr_p40}
\end{figure*}

In this section, we complement the parameter exploration done in Sect.\,\ref{sect:results_2} by exploring the same diagrams at different $\phi_b$. We conclude that the trends observed at $\phi_b=-30$\,deg (Fig.\,\ref{fig:corr_p30}) are maintained when looking at other azimuths.
The only noticeable change is the steepness of the \dR{ }and \dphi{ }panels according to the distance to the major axis of the bar. In Fig.\,\ref{fig:corr_p20} ($\phi_b=-20$\,deg) the contour lines are close to each other, indicating a rapid change of the properies. Opposite to that, in Fig.\,\ref{fig:corr_p40} ($\phi_b=-40$\,deg), the contour lines are more sepparated, indicating a solution which is more stable.

\section{Model vs Model degeneracy}\label{sect:degeneracy_appendix}

\begin{figure}
    \centering
    \includegraphics[width=0.49\textwidth]{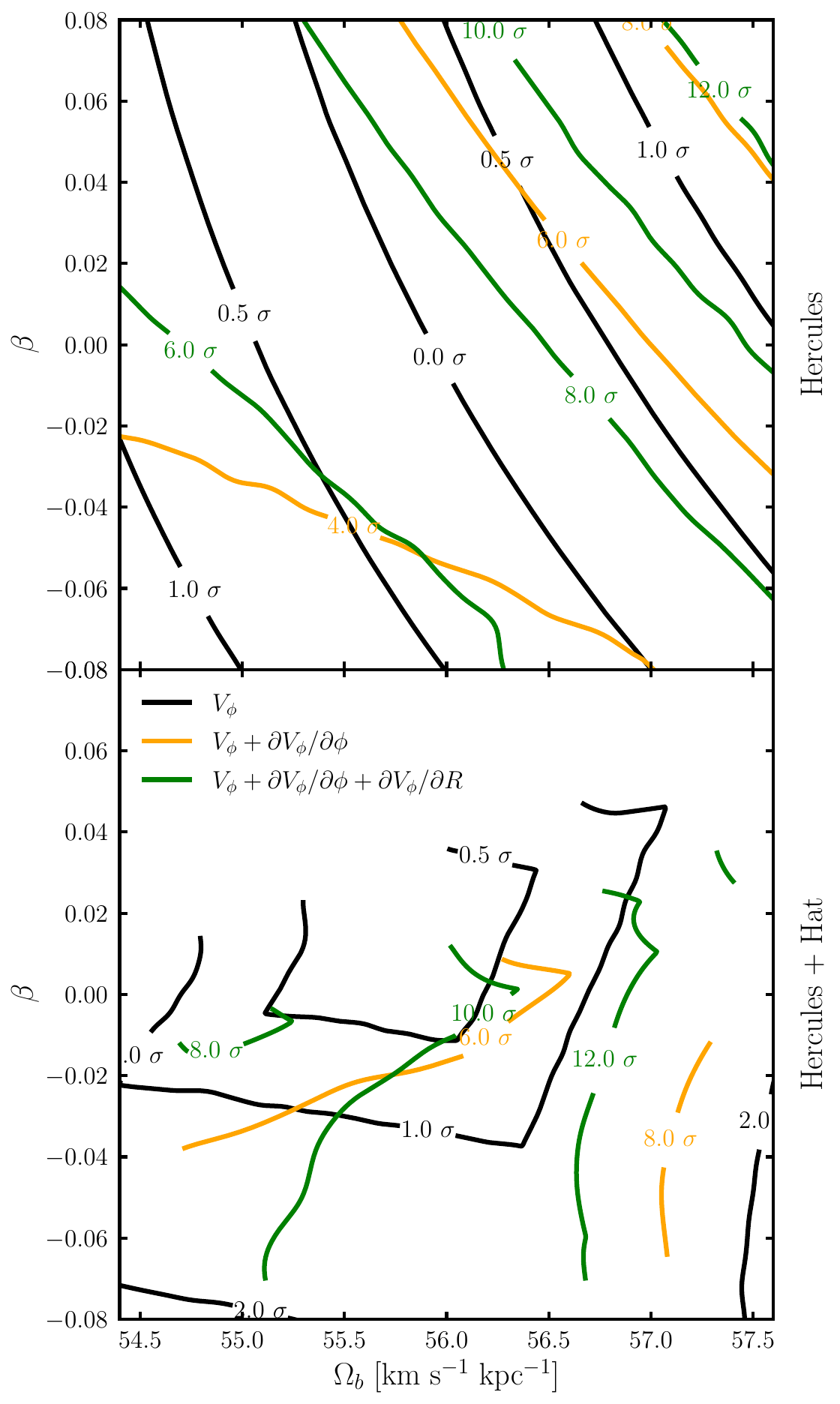}
    \caption{Differences between the fiducial SBM and the galaxy models around the FBM for Hercules \textit{(top)} and Hercules + Hat \textit{(bottom)}. Black contours: Disagreement considering only $V_\phi$ measurements. Orange contours: Disagreement considering $V_\phi$ and \dphi. Green contours: Disagreement considering $V_\phi$, \dphi{} and \dR.}
    \label{fig:fitting_SMBvsFBM}
\end{figure}

\begin{figure}
    \centering
    \includegraphics[width=0.49\textwidth]{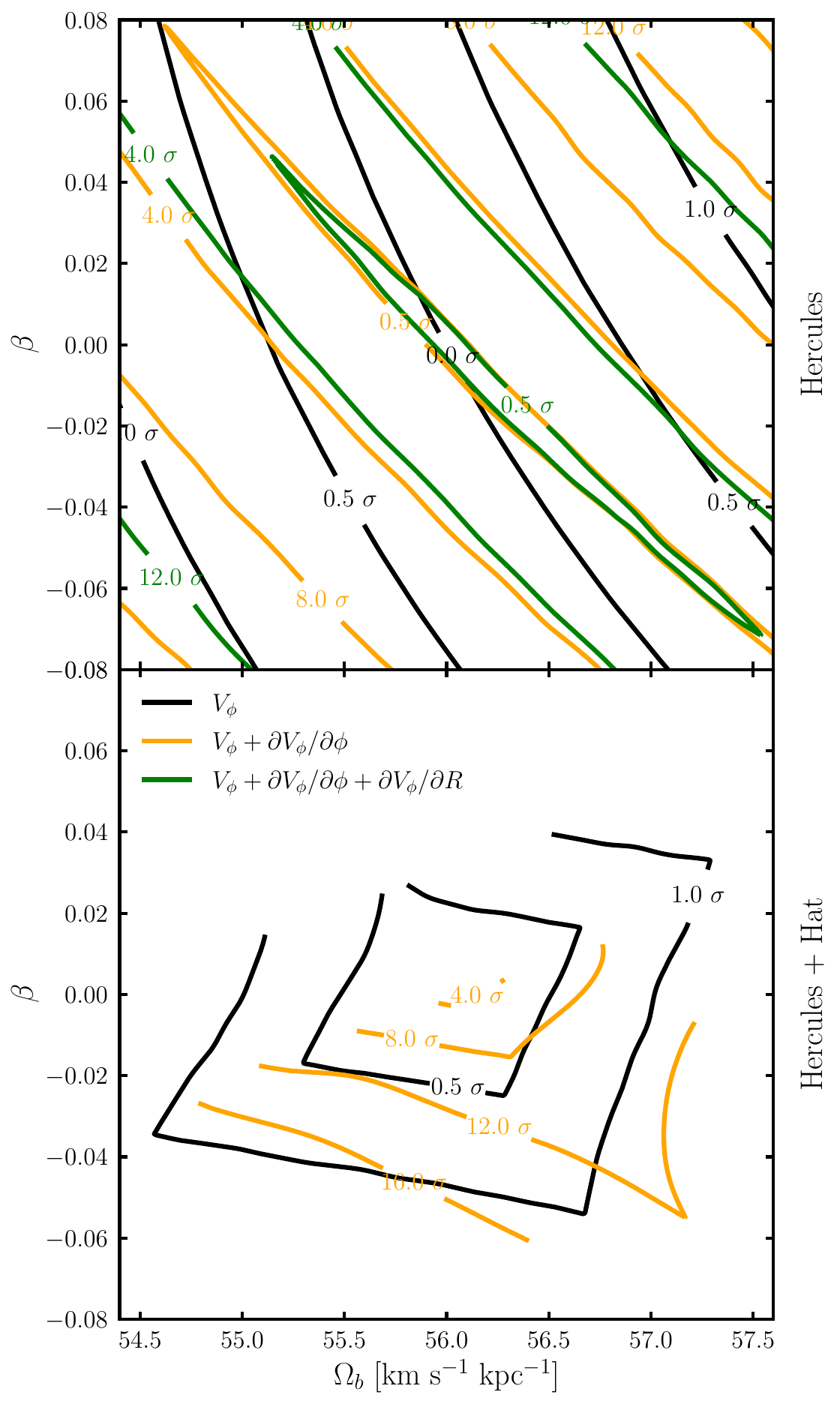}
    \caption{Similar to Fig.~\ref{fig:fitting_SMBvsFBM} but showing the differences between the fiducial FBM and the galaxy models around it when using only the Hercules-like overdensity (top) or Hercules- and Hat-like (bottom).}
    \label{fig:fitting}
\end{figure}


The computation of the Galactic models links a set of \emph{parameters} ($\Omega_b, \beta, \phi_b...$) to a set of \emph{measurables} ($V_R$, $V_\phi$, \dR, \dphi).
For a set of \emph{target measurables} (e.g. \emph{Gaia} DR3 measurables), a set of parameters will be \emph{optimal} if the measurables it produces match the target measurables. Given a set of parameters, we define their \emph{statistical difference} ($\sigma$, dimensionless) with respect to a target as the difference between the measurables in units of the measure error in the target.
Notice that in this work we have not been able to obtain an optimal set of parameters to reproduce the target \emph{Gaia} DR3 measurables. In this section we will use the fiducial models as targets.

Lets suppose that our target measurables are the ones from the fiducial SBM (Tab.\,\ref{tab:slopes}). In this case, by definition the fiducial SBM will be an optimal set of parameters. To test the degeneracies, we want to check if there are other sets of parameters that also reproduce the SBM measurables.
This is what we explore in Figure \ref{fig:fitting_SMBvsFBM}, where we show the statistical differences between the target measurables (assuming those of the model SBM) and the models around the FBM using various combinations of measurables. The top panel exclusively utilizes the measurables corresponding to the Hercules-like overdensity, while the bottom panel incorporates measurables for both Hercules and Hat. The black contours represent the statistical differences between the model and the target when considering only $V_\phi$. When restricted to Hercules and its local value of $V_\phi$ alone (black contour, top panel), we observe a substantial region of parameter space where the solution is degenerate, with multiple combinations of $\Omega_b$ and $\beta$ from the FBM models yielding the same $V_\phi$. However, as we introduce additional measurables, such as \dphi{ }(orange contour) and \dR{ }(green contour), the statistical difference becomes more pronounced, scaling to over $4\sigma$. When we further include a second structure like Hat (bottom panel), the degeneracy in the solution can be partially resolved using only $V_\phi$. Once gradients are factored in, the distinction between models is once again emphasized.

To complement this study, we repeat the analysis assuming that the target measurables are the ones of the fiducial FBM (Table \ref{tab:slopes}), and study the statistical difference with the models around it (Fig.\,\ref{fig:fitting}). The goal in this exercise is to characterise the local degeneracy. The results are very similar to the ones in Fig.\,\ref{fig:fitting_SMBvsFBM}: we need to include the gradients to obtain a statistical difference.

This analysis suggests that the best combination of measurables to break the degeneracy is to consider both Hercules and Hat (bottom panel). Using only $V_\phi$ of both structures we already see that we are able to constrain a small region of $\Omega_b$ and $\beta$. In addition, including the azimuthal gradient brings the statistical significance over $8\sigma$ in almost the entire parameter space.

\end{appendix}

\end{document}